%% file: IoTtrafficDrift.tex
\documentclass[journal,10pt]{IEEEtran}
\usepackage{amsmath,amsfonts}
\usepackage{algorithmic}
\usepackage{algorithm}
\usepackage{array}
\usepackage{subcaption}
\usepackage{textcomp}
\usepackage{stfloats}
\usepackage{url}
\usepackage{verbatim}
\usepackage{hyperref}
\usepackage{xcolor}
\usepackage{graphicx}
\usepackage{cite}
\usepackage{balance}
\newcommand{\ie}{\textit{i.e.,}\ }
\newcommand{\eg}{\textit{e.g.,}\ }
\newcommand{\myverbB}{\fontsize{9}{16}\usefont{OT1}{lmtt}{b}{n}\noindent }
\newcommand{\myverb}{\fontsize{7}{16}\usefont{OT1}{lmtt}{b}{n}\noindent }

\begin{document}

\title{
{\scriptsize\textnormal{This work has been submitted to the IEEE for possible publication. Copyright may be transferred without notice, after which this version may no longer be accessible.}}
\\[1.5ex]
Maintaining IoT Device Identification under Concept Drift  via Budget-Aware Traffic Labeling
}

\author{Shayan~Azizi,
        Norihiro~Okui,
        Masataka~Nakahara,
        Ayumu~Kubota,
        \\Gustavo~Batista,        and~Hassan~Habibi~Gharakaheili
        \thanks{S.~Azizi and H.~Habibi~Gharakaheili are with the School of Electrical Engineering and Telecommunications, University of New South Wales, Sydney, NSW 2052, Australia (e-mails: \{s.azizi, h.habibi\}@unsw.edu.au).}
        \thanks{G.~Batista is with the School of Computer Science and Engineering, University of New South Wales, Sydney, NSW 2052, Australia (e-mail: g.batista@unsw.edu.au).}
        \thanks{N.~Okui, M.~Nakahara and A.~Kubota, are with KDDI Research, Japan (e-mails: \{no-okui, ms-nakahara, ay-kubota\}@kddi.com).
        }
    }

\maketitle

\input{sections/0abstract}


\input{sections/1intro}

\vspace{5mm}
\input{sections/2prior}

\input{sections/3dataset}

\input{sections/4drift}

\input{sections/5distribution}

\input{sections/6evaluation}


\input{sections/7limitations}

\input{sections/8conclusion}

\bibliographystyle{IEEEtran}
\bibliography{IoTtrafficDrift}

\appendix

\input{sections/apdx}

\end{document}

%% file: sections/0abstract.tex
\begin{abstract}

    Identification of IoT device types from passive traffic is increasingly used for security management in enterprise and ISP networks. However, the performance of machine learning-based classifiers gradually degrades under concept drift as device behavior evolves. Therefore, maintaining classification performance requires periodic retraining with newly labeled deployment traffic. The operational challenge is determining how much and which deployment traffic instances to label for maintaining classification performance. We show that these two decisions should be treated separately. While retraining solely on instances selected by a drift detector is prone to systematically overlooking parts of the emerging behavioral space, uniformly sampled deployment traffic captures more representative behavioral changes. Instead, drift detection is more effective at determining the amount of deployment traffic that should be labeled. We make three contributions. (1) We conduct a two-year longitudinal study of IoT traffic and characterize how behavioral evolution manifests across device classes and how retraining with newly labeled traffic restores classification performance. (2) We develop a conformity-based drift detector that captures class-conditional behavioral models directly from raw traffic features and provides feature-level explanations of behavioral evolution. (3) We demonstrate that adjusting the traffic labeling rate according to the observed behavioral evolution, combined with uniform traffic sampling, maintains classifier performance more effectively than detector-guided sample selection and is beneficial to managing the traffic labeling effort. We further show that this strategy performs comparably to confidence-guided adaptation while providing feature-level explanations. Our evaluation uses 3.8 million IPFIX flow records collected from 21 IoT types over more than 2 years.    
\end{abstract}

%% file: sections/1intro.tex
\vspace{-3mm}
\section{Introduction}

The rapid growth of Internet-of-Things (IoT) devices in residential and enterprise networks has made visibility into connected devices essential for effective security management. Many IoT devices run lightweight firmware \cite{Williams:ElsevierIoT2022}, receive limited long-term vendor support \cite{UlHaq:DiscoverIoT23}, and suffer from authentication weaknesses \cite{Alrawi:SP19,JayAnand:DAISNAC21}, making them particularly susceptible to cyber compromise. These risks range from the possibility of device seizure, which can be exploited for large-scale volumetric attacks \cite{Kumari:ElsevierCS23,Mirai,Feamster:SPW18}, to user behavior tracking and private data exfiltration \cite{Choffnes:PETS20,Selcuk:USENIXSec18,Iqbal:IMC23,Choffnes:IMC19,Feamster:NDSS16,Sadeghi:WiSec20,Feamster:NDSS26}. 
These threats are particularly important for Internet Service Providers (ISPs), which provide connectivity to millions of residential and enterprise subscribers.  
Cyber incidents originating from vulnerable IoT devices can expose ISPs to malicious traffic, service disruption, and resource exhaustion \cite{Arman:Deterministic}. 
Enterprise network operators face similar challenges and must continuously discover and monitor these assets to maintain network security.

Mitigating the cyber threats posed by IoT devices requires identifying the devices so that appropriate security policies can be applied. To support this, the community has curated vulnerability inventories \cite{NVD,CVE} and traffic allowlists \cite{mandalari:PETS21,Ayyoob:TDSC20,Feamster:EuroSP24}. Leveraging these resources requires operators to first  identify IoT devices from network traffic so that device-specific policies can be applied. For example, ISPs are highly interested in scalable techniques that infer connected devices from passively collected
 traffic at their network vantage points \cite{Arman:Deterministic,IMC2020Haystack}, enabling subscribers to be informed about vulnerable devices and appropriate mitigation strategies. Accordingly, mainstream IoT identification systems rely on Machine Learning (ML) classifiers trained on network traffic collected at different granularities \cite{Sivanathan:TMC19,Marchal:JSAC19,Miettinen:ICDCS17,DEFT,Arman:Deterministic,Ahmed:PETS22}. As these classifiers become operational components of enterprise and ISP security infrastructures, maintaining their accuracy throughout long-term deployment becomes an operational requirement. 

Once deployed, IoT device classifiers cannot be expected to remain accurate indefinitely. As device behaviors evolve over time because of firmware updates, software changes, and changing operating environments, the underlying traffic patterns also evolve. From the perspective of a deployed classifier, this behavioral evolution manifests as \textit{concept drift}, a gradual degradation in classifier performance. Therefore, maintaining classifier accuracy becomes an operational problem instead of a one-time training problem. 
Existing work has reported classifier degradation under concept drift \cite{Kolcun:TMA2021,Azizi:Sigcomm24,Maali:NDSS25,Arman:IoTJ23}, investigated pseudo-labeling using foundation models \cite{Shahbaz:OSDI24}, or proposed early mitigation strategies with limited efficacy\cite{Arman:IoTJ23}. However, none addressed the practical problem of maintaining deployed IoT traffic classifiers over long periods under limited traffic-labeling budgets.

Since labeling deployment traffic is operationally expensive \cite{Guerra:COSE22}, operators can label only a limited fraction of deployment traffic. This introduces two  practical questions: how much deployment traffic should be labeled, and which deployment instances should be selected. Existing work has largely focused on the latter by using  drift detectors to identify informative deployment instances for labeling. In this paper, we challenge this assumption. We show that the detector-selected instances are biased toward only a subset of the emerging behavioral patterns, causing retrained classifiers to overlook other aspects of behavioral evolution. In contrast, uniformly sampled deployment traffic captures a broader range of evolving behaviors and consistently provides more effective classifier maintenance.

Having established that uniformly sampled deployment traffic is preferable for retraining, the remaining question is how much traffic should be labeled at each stage of deployment.  
Since the degree of behavioral evolution varies over time, a fixed labeling rate is unlikely to utilize a limited operational labeling budget efficiently throughout the lifetime of the classifier. This calls for a different role of drift detection.  
Existing approaches estimate drift either from classifier confidence scores \cite{Jordaney:USENIXSecurity17,transcendent,Han:NDSS23,Zhang:INFOCOM20,Jorgensen:TAI24} or from learned feature embeddings \cite{CADE:USENIXSecurity21,Chen:USENIXSec23}. 
Confidence-based methods provide limited visibility into underlying traffic changes \cite{NeurIPS19}, while embedding-based approaches increase computational complexity and obscure feature-level interpretation. Neither provides an interpretable behavioral signal for guiding budget-aware classifier maintenance.

Accordingly, we develop a budget-aware adaptation framework that separates the two decisions introduced above: determining how much deployment traffic should be labeled and deciding which deployment instances should be selected. Our conformity-based detector measures behavioral evolution by identifying feature-level non-conformity with respect to class-conditional traffic distributions. It also provides interpretable feature-level explanations for the detected behavioral changes. The resulting behavioral signal guides the traffic labeling rate, while deployment traffic itself is sampled uniformly  for retraining.  
Newly labeled traffic is then incorporated into classifier retraining, and the detector is updated accordingly, enabling the continuous maintenance of IoT traffic classifiers as device behaviors evolve over time.

Our specific contributions are as follows. 
\textbf{(1)} We conduct a longitudinal study (\S\ref{sec:conceptDrift}) of IoT traffic spanning more than two years and 21 real device types, and characterize how behavioral evolution manifests across device classes and how it induces concept drift in deployed classifiers. We further demonstrate that incorporating freshly labeled deployment traffic enables effective classifier maintenance.
\textbf{(2)} We develop an instance-level conformity-based detector (\S\ref{sec:detection}) that models class-conditional feature behavior and provides feature-level explanations of behavioral evolution. The resulting behavioral signal is later used to guide the utilization of a limited labeling budget throughout deployment. 
\textbf{(3)} We demonstrate that adjusting the traffic labeling rate according to the observed behavioral evolution effectively maintains classifier performance over two years of deployment (\S\ref{sec:adaptEval}). Compared with different confidence-based adaptation strategies, the proposed approach achieves comparable or better performance at a lower labeling cost, while additionally identifying the traffic features associated with behavioral evolution.

%% file: sections/2prior.tex
\section{Related Work}

\textbf{Modeling network behavior of IoT devices:} 
A large body of work has shown that IoT devices exhibit distinctive network behaviors that can be modeled from network traffic for device identification and anomaly detection \cite{Sivanathan:TMC19,Marchal:JSAC19,Miettinen:ICDCS17,DEFT,Kolcun:TMA2021,Kolcun:arXiv20,Ahmed:PETS22,Arman:Deterministic,23TIOT,25tnse}. Existing approaches use packet-level features (\eg protocol fingerprints, domain names, TLS metadata, and TCP options), flow-level statistics (\eg packet and byte counts, timing characteristics, and periodicity), or combinations of both to characterize IoT behaviors. 
These behavioral models have enabled accurate device identification and behavioral monitoring in enterprise, residential, and ISP environments. However, they generally assume that the learned behavior remains representative after deployment or that updated labeled traffic is readily available for retraining. Our work instead investigates how these behavioral models evolve over time and how they can be maintained efficiently under limited traffic-labeling resources.

\textbf{Concept drift and drift detection in networks and systems:} Behavioral evolution in IoT traffic over time or across deployment environments has been observed in several recent studies, often through its impact on classifier performance  \cite{Maali:NDSS25,Azizi:Sigcomm24,Arman:IoTJ23,Ahmed:PETS22,Kolcun:TMA2021,Kolcun:arXiv20,Yang:GLOBECOM21,Ghorbani:IoTJ24}. 
However, most studies report this degradation as an observed effect, offering limited insight into the underlying behavioral evolution or into how classifiers should be maintained over long-term deployment. 
Some works correlate performance degradation with changes in the distribution of the most important features identified by the classifier \cite{Maali:NDSS25} and \cite{Azizi:Sigcomm24}. Others use accuracy degradation as a signal to trigger model updates \cite{Yang:GLOBECOM21}. However, such approaches often assume access to ground-truth labels, which is unrealistic in practical deployments. Similarly, \cite{carnier:arxiv25} assumes an online learning setting where all deployment traffic is labeled, but does not address selective labeling. 

Alternative approaches attempt to mitigate drift without explicit detection. For example, work in \cite{Arman:IoTJ23}  selects from a pool of historical models  
based on the distributional similarity of prediction scores to the training scores of each model. However, such methods cannot handle the 
previously unseen drift patterns, nor do they identify drift instances for targeted adaptations. 
Other studies examine variability in IoT traffic behavior, such as changes in cloud endpoints over time and across deployments \cite{Feamster:EuroSP24}. Meanwhile, \cite{Azizi:CPSIoTSec25} focuses on label shift, in which class prevalence changes while per-class distributions remain stable, which is outside the scope of our work. 

Beyond IoT traffic classification, concept drift has also been recognized as a general problem in networking and systems applications. 
Some works demonstrate classifier degradation under evolving traffic patterns \cite{Malekghaini:CompNet2023}, while others explore detection or adaptation under different assumptions. 
For example, work in \cite{Feamster:CONEXT23} detects drift in time-series forecasting (\ie a regression problem instead of classification) by monitoring prediction error. 
Other works employ auxiliary models to generate pseudo-labels, enabling adaptation in the absence of ground truth \cite{Shahbaz:OSDI24,Andresini:AISec21}, but they introduce additional model complexity and potential error propagation. 
The authors of \cite{Xavier:APNet24} propose an in-network drift detection mechanism, but their method still relies on ground-truth labels, making it less suitable for real-world scenarios where labeled data is scarce or unavailable. 

Model selection approaches, such as \cite{Arzani:MLSys22}, choose among multiple pre-trained models, but they are not designed to handle previously unseen behavioral patterns. 
Transfer learning \cite{Malekghaini:IFIPNetworking23} and domain adaptation methods \cite{Bertino:NDSS25,APIGraph:CCS20} aim to improve the model robustness to changes but do not address instance-level drift detection.

Beyond these adaptation strategies, a large body of work focuses specifically on detecting concept drift.
Existing drift detection methods generally fall into three categories.
The majority detect drift from classifier output confidence \cite{Jorgensen:TAI24,Wang:TIFS25,Zhang:INFOCOM20,Jordaney:USENIXSecurity17,transcendent, Han:NDSS23, EMA:arxiv26}, which provides limited visibility into the underlying behavioral changes. 
Another line of work detects drift in learned feature embeddings  \cite{CADE:USENIXSecurity21, Chen:USENIXSec23}. Although these methods improve robustness to complex traffic patterns, they rely on latent representations that increase computational complexity and obscure feature-level interpretation.  
A smaller body of work detects drift directly from raw input features \cite{Qahtan:KDD15, Gustavo:KDD16, SanjayRanka:KDD07}, but these methods focus on aggregate distribution shifts rather than instance-level detection and have not been evaluated for long-term IoT traffic classification.
A related line of work models feature distributions for downstream tasks, such as noisy-label correction in anomaly detection \cite{Qing:NDSS24}. Unlike our marginal modeling approach, these methods rely on deep generative models, such as MADE \cite{MADE}, to model the joint feature distribution, making them computationally expensive and data-hungry. Hence, these traffic modeling approaches are less suitable for IoT traffic settings where limited samples are available per class.

Collectively, these studies demonstrate the importance of detecting concept drift but offer limited insight into the behavioral evolution that drives classifier degradation and into how that information should inform long-term classifier maintenance under limited traffic-labeling budgets. These observations motivate approaches that directly model behavioral evolution from raw traffic features while providing interpretable drift detection and efficient classifier maintenance.

\textbf{Drift adaptation:} 
Maintaining a classifier under concept drift requires not only detecting behavioral changes but also updating the classifier using newly labeled deployment traffic. Existing research has chiefly focused on the mechanics of model adaptation once labeled data are available rather than on how scarce labeling resources should be allocated during deployment.
Numerous approaches have been proposed for model adaptation \cite{PMLR:adaptation}. These approaches include model re-training \cite{Bifet:ADWIN, Bifet:STUDD, CADE:USENIXSecurity21}, ensemble techniques \cite{JMLR:kolter, SEA:KDD01}, gradient-based updating \cite{Chen:USENIXSec23, Zhang:CVPR20} or warm-start re-training according to similarity to previous data \cite{EMA:arxiv26}. While the majority of adaptation approaches prioritize the most recently observed data, either explicitly or implicitly, to improve responsiveness to concept drift, doing so may increase susceptibility to catastrophic forgetting \cite{Feamster:CONEXT23, catastrophic_forgetting}. 
Throughout this paper, we adopt cumulative retraining during the evaluation to isolate the effect of traffic-labeling strategies from the choice of adaptation algorithm. A comprehensive comparison of alternative adaptation mechanisms is beyond the scope of this work.

%% file: sections/3dataset.tex
\begin{table}[t]
\caption{The 21 IoT devices studied.}
\label{tab:devices}
\centering
\renewcommand{\arraystretch}{1.1}
\begin{tabular}{p{4.7cm} p{2cm}}
\hline
\textbf{Device Make-and-Model} & \textbf{Category} \\
\hline
Amazon Echo Show 5  & Voice assistant \\
Amazon Echo Gen 2  & Voice assistant \\
Google Nest Mini  & Voice assistant \\
Amazon Echo Dot with Clock  & Voice assistant \\
Google Home  & Voice assistant \\
Sony Smart Speaker & Voice assistant \\
Line Clova Wave & Voice assistant \\
Apple Homepod  & Voice assistant \\
Google Chromecast  & Streamer \\
Amazon Fire TV Stick 4K & Streamer \\
I-O DATA QWatch Camera & IP Camera \\
Wansview Q5 & IP Camera \\
Panasonic Doorphone & Doorbell \\
TP-Link Kasa Smart LED Bulb & Smart LED \\
TP-Link Kasa Smart Plug & Smart plug \\
TP-Link Kasa Pro & Smart power strip \\
Nature Remo & Remote control \\
Philips Hue Bridge & Smart home hub \\
Panasonic Home Unit & Smart home hub \\
JVCKENWOOD Hub  & Smart home hub \\
Withings Nokia Body & Smart scale \\
\hline
\end{tabular}
\end{table}

\section{IoT Traffic Data}
We use a longitudinal IoT dataset collected over more than 2 years from 21 device types (make-and-models) in a controlled lab testbed. The devices span multiple categories, including smart speakers, cameras, streaming devices, smart plugs, smart hubs, and home automation equipment. Table~\ref{tab:devices} summarizes the inventory of device types.

In this study, we focus on TCP/443 (presumably HTTPS) traffic due to its widespread adoption across IoT devices and the rich statistical semantics captured by flow-level features. Network flows are represented as IPFIX records exported from raw PCAP files using YAF \cite{YAFdoc}. Each flow is described using 22 statistical features, including packet count, byte count, packet size, and timing statistics measured in both the forward and reverse directions. Table~\ref{tab:features} summarizes the traffic features.

In certain cases, late-arriving packets for an already-exported flow can result in additional redundant flow records with limited or noisy semantics. We filter out such redundancies based on temporal proximity among flows that share the same 5-tuple. 

We use traffic collected between 1 Oct 2021 and 31 Dec 2021 for training and validation, and data between 1 Jan 2022 and 31 Dec 2023 for evaluation. The resulting dataset consists of approximately 260K training flows, 113K validation flows, and 3.38M deployment-phase flows spanning two years of operation. Our data collection and analysis raise no ethical concerns.

\begin{table}[t]
\caption{Traffic features in IPFIX flow records.}
\label{tab:features}
\centering
\resizebox{\columnwidth}{!}{
    \renewcommand{\arraystretch}{1.2}
    \begin{tabular}{p{4.2cm} p{4.2cm} p{3.5cm}}
    \hline
    \textbf{Feature Name} & \textbf{Description} & \textbf{Abbr. Name (reverse)} \\
    \hline
    {\myverb{packetTotalCount}} & \# of packets. & {\myverb{pktTotCnt (rPktTotCnt)}} \\
    {\myverb{octetTotalCount}} & \# of bytes (header and payload). & {\myverb{octTotCnt (rOctTotCnt)}} \\
    {\myverb{smallPacketCount}} & \# of packets with less than 60 Bytes \\&of payload. & {\myverb{sPktCnt (rSPktCnt)}} \\
    {\myverb{largePacketCount}} & \# of packets with at least 220 Bytes \\&of payload. & {\myverb{lPktCnt (rLPktCnt)}} \\
    {\myverb{nonEmptyPacketCount}} & \# packets with payload. & {\myverb{nonEmptPktCnt (rNonEmptPktCnt)}} \\
    {\myverb{dataByteCount}} & \# of payload bytes. & {\myverb{datByteCnt (rDatByteCnt)}} \\
    {\myverb{averageInterarrivalTime}} & Average time (ms) between packets. & {\myverb{avgIntArrTime (rAvgIntArrTime)}} \\
    {\myverb{firstNonEmptyPacketSize}} & \# of bytes in the payload of the \\&1st non-empty packet. & {\myverb{1stNonEmptPktSz (r1stNonEmptPktSz)}} \\
    {\myverb{maxPacketSize}} & largest payload size (bytes). & {\myverb{maxPktSz (rMaxPktSz)}} \\
    {\myverb{standardDeviationPayloadLength}} & std. deviation of payload size for \\&first 10 non-empty packets. & {\myverb{stdDevPLlen (rStdDevPLlen)}} \\
    {\myverb{standardDeviationInterarrivalTime}} & std. deviation of time (ms) \\&between first 10 non-empty packets. & {\myverb{stdDevIntArrTime (rStdDevIntArrTime)}} \\
    \hline
    \end{tabular}
    }
\end{table}

%% file: sections/4drift.tex
\section{Behavioral Evolution of IoT Traffic \\under Long-Term Deployment}\label{sec:conceptDrift}
In this section, we investigate how behavioral evolution during long-term deployment affects machine learning-based IoT traffic classification. We first establish a baseline IoT traffic classifier and evaluate its performance over a two-year deployment period without model updates. We then analyze the observed performance changes to characterize the underlying behavioral evolution, showing that it is continuous, heterogeneous across device classes, and not necessarily monotonic. Finally, we demonstrate that classifier performance can be effectively restored by incorporating freshly labeled deployment traffic, establishing that long-term classifier maintenance is feasible and motivating the search for practical maintenance strategies under limited labeling resources.

\subsection{IoT Traffic Classification Setup}
\label{subsec:classifier}
To study the impact of behavioral evolution on IoT traffic classification, we first establish a baseline traffic classifier.  
We use a Random Forest classifier, which has demonstrated strong performance in network traffic classification tasks \cite{Ahmed:PETS22,Yifan:IMC24,Kolcun:TMA2021}. The classifier is trained on 70\% of the traffic flows collected between 1 Oct 2021 and 31 Dec 2021, with the remaining 30\% reserved for validation. The split is performed randomly rather than chronologically so that both partitions are drawn from the same underlying distribution. 

The classifier hyper-parameters are selected using 10-fold cross-validation on the training data. The resulting model consists of 100 decision trees with a maximum depth of 30 and a minimum of 5 samples per split. Evaluated on the validation set, the classifier achieves a macro-averaged F1-score of $0.974$. This classifier serves as the baseline model for the remainder of the paper to analyze behavioral evolution and evaluate maintenance strategies under long-term deployment.

\subsection{Behavioral Evolution under Long-Term Deployment}
\begin{figure}[!t]
  \centering
  \includegraphics[width=0.995\linewidth]{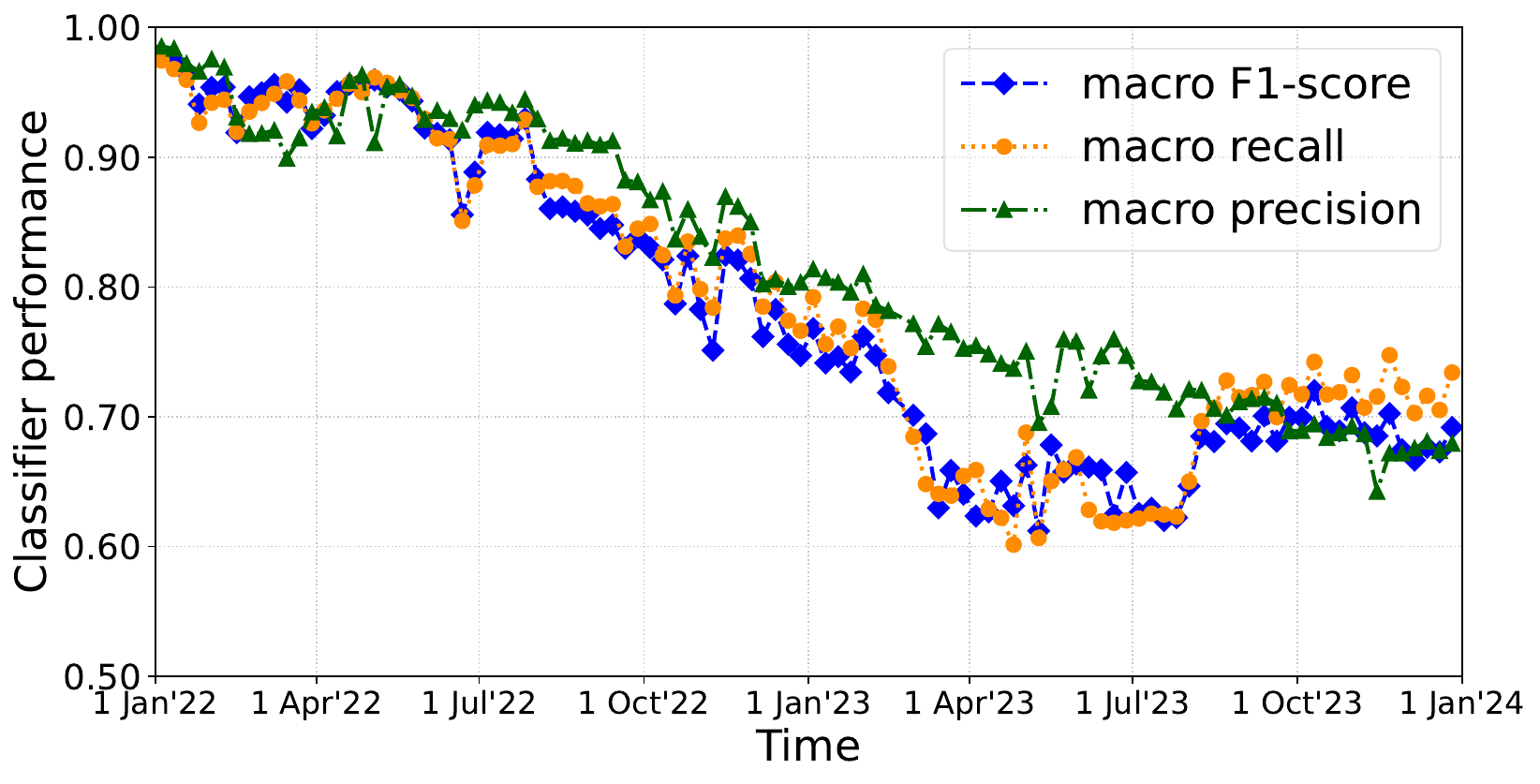} 
  \vspace{-5mm}  
  \caption{Behavioral evolution during long-term deployment progressively degrades classifier performance. The partial restoration observed in late 2023 suggests that device behaviors may evolve non-monotonically away from the training distribution.}\label{fig:performanceDrop}
  \vspace{-3mm}
\end{figure}

After establishing the baseline classifier, we deploy it on previously unseen traffic collected continuously from January 2022 onward without updating the model. To assess the performance of the classifier, we divide the deployment traffic into consecutive 7-day windows from the beginning of 2022 until the end of 2023. Fig. \ref{fig:performanceDrop} summarizes the classifier's performance on a weekly basis over the subsequent two years. A progressive decline in the macro F1-score is observed beginning around mid-2022, indicating that the traffic patterns learned during training gradually become less representative of the deployment traffic. 
From the perspective of the deployed classifier, this behavioral evolution manifests as \textit{concept drift}, progressively degrading classification performance.
Interestingly, the degradation is not monotonic. Instead, classifier performance exhibits a partial restoration during the second half of 2023, suggesting that IoT device behaviors evolve dynamically rather than continuously drifting away from the training distribution.

\begin{figure*}[!t]
\centering
\begin{subfigure}{0.48\textwidth}
    \centering
    \includegraphics[width=\linewidth]{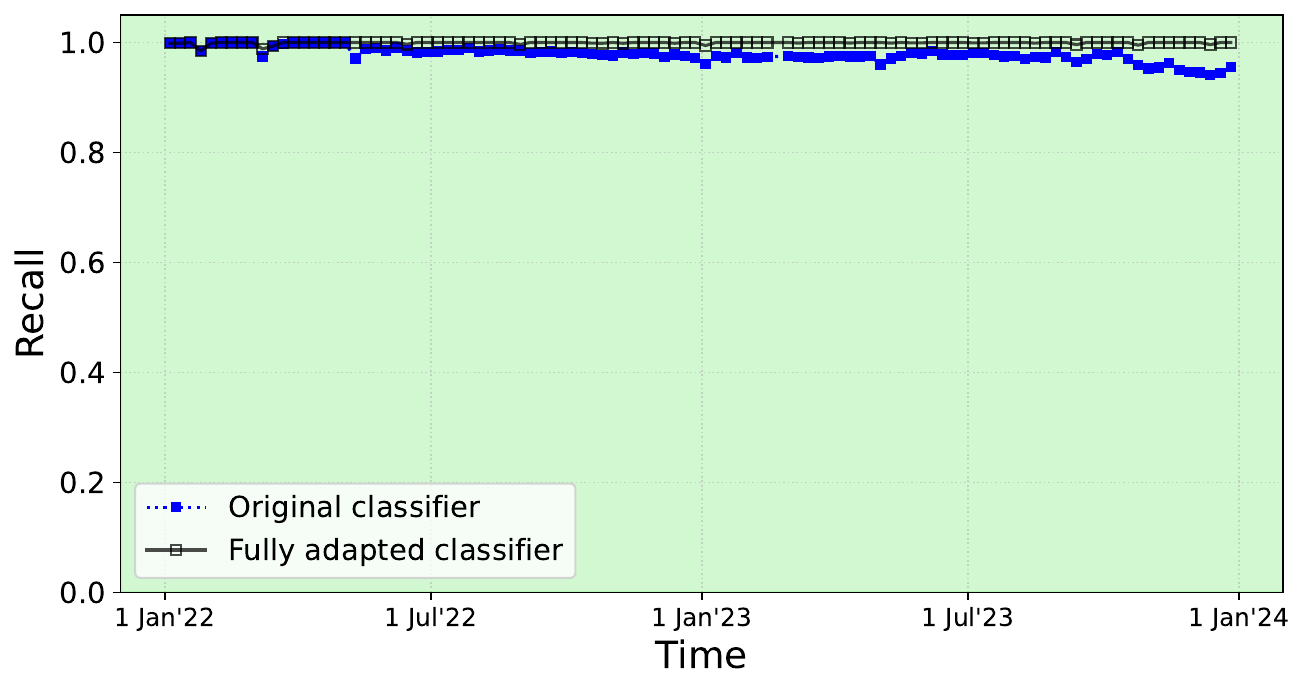}
    \caption{Persistent behavior in Philips Hue Bridge.}
    \label{fig:persistentPerf}
\end{subfigure}
\hfill
\begin{subfigure}{0.48\textwidth}
    \centering
    \includegraphics[width=\linewidth]{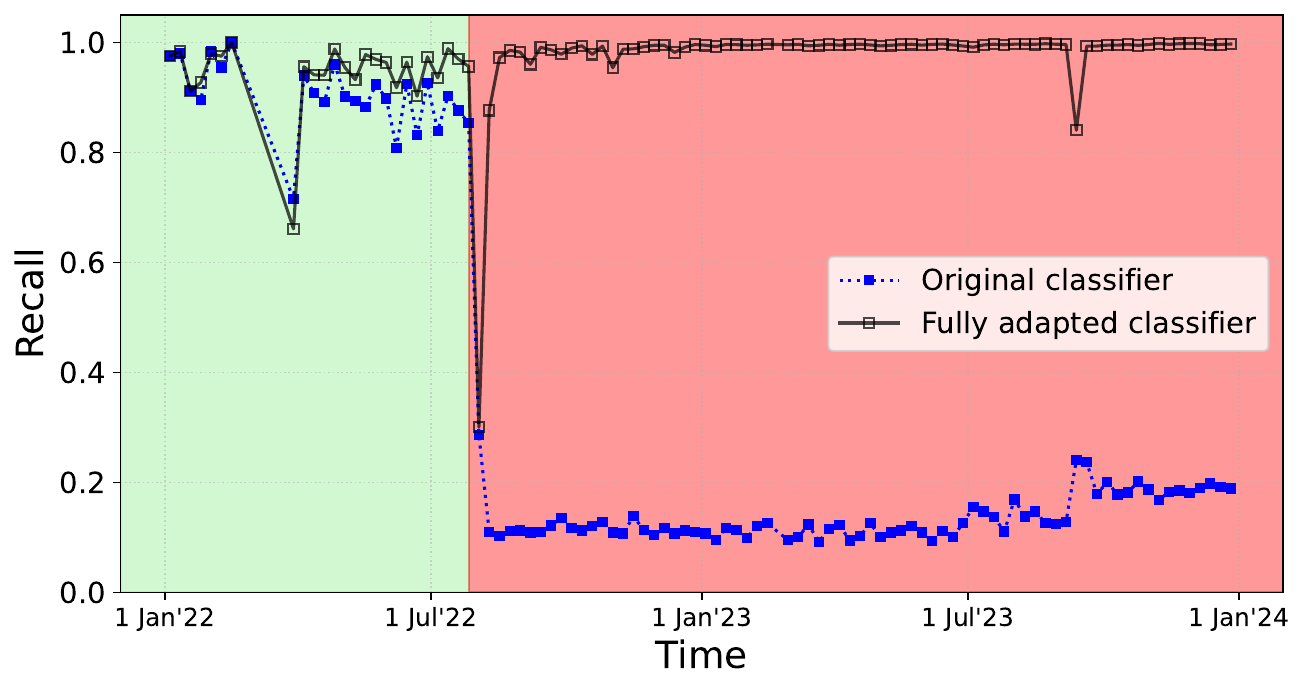}
    \caption{Behavioral decay in Google Nest Mini.}
    \label{fig:decayingPerf}
\end{subfigure}

\vspace{0.5cm}

\begin{subfigure}{0.48\textwidth}
    \centering
    \includegraphics[width=\linewidth]{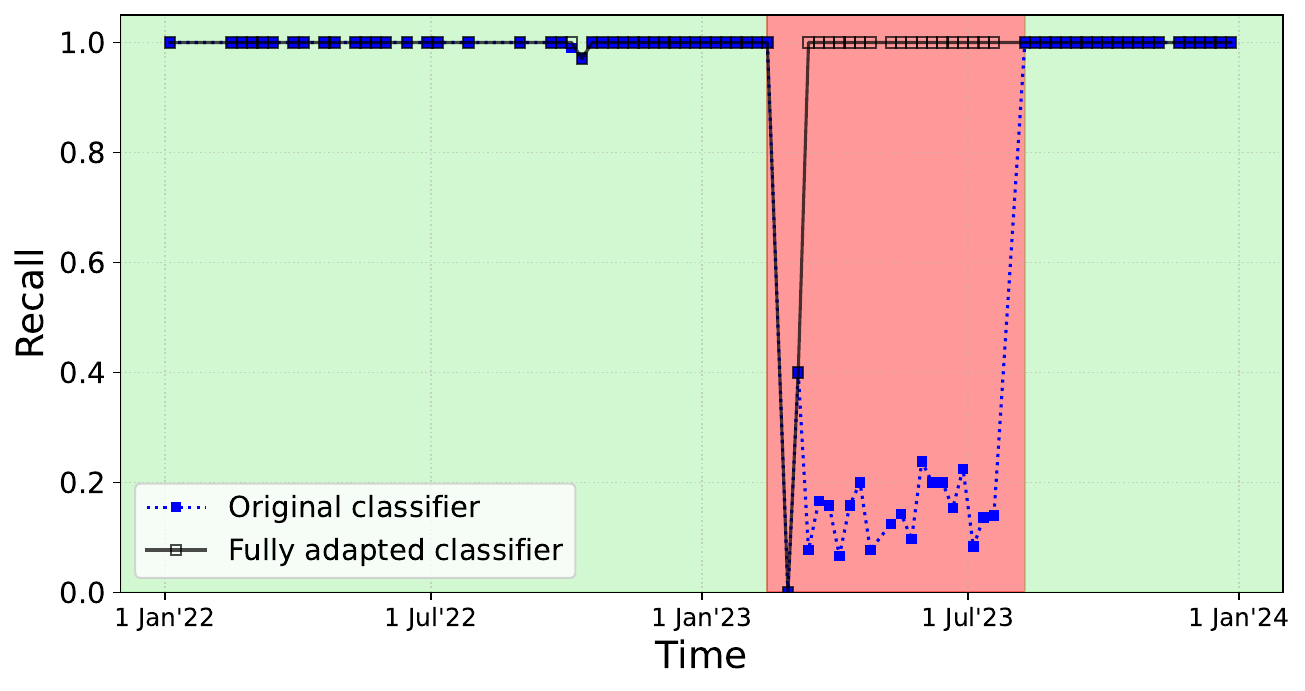}
    \caption{Behavioral restoration in JVC Hub.}
    \label{fig:restoringPerf}
\end{subfigure}
\hfill
\begin{subfigure}{0.48\textwidth}
    \centering
    \includegraphics[width=\linewidth]{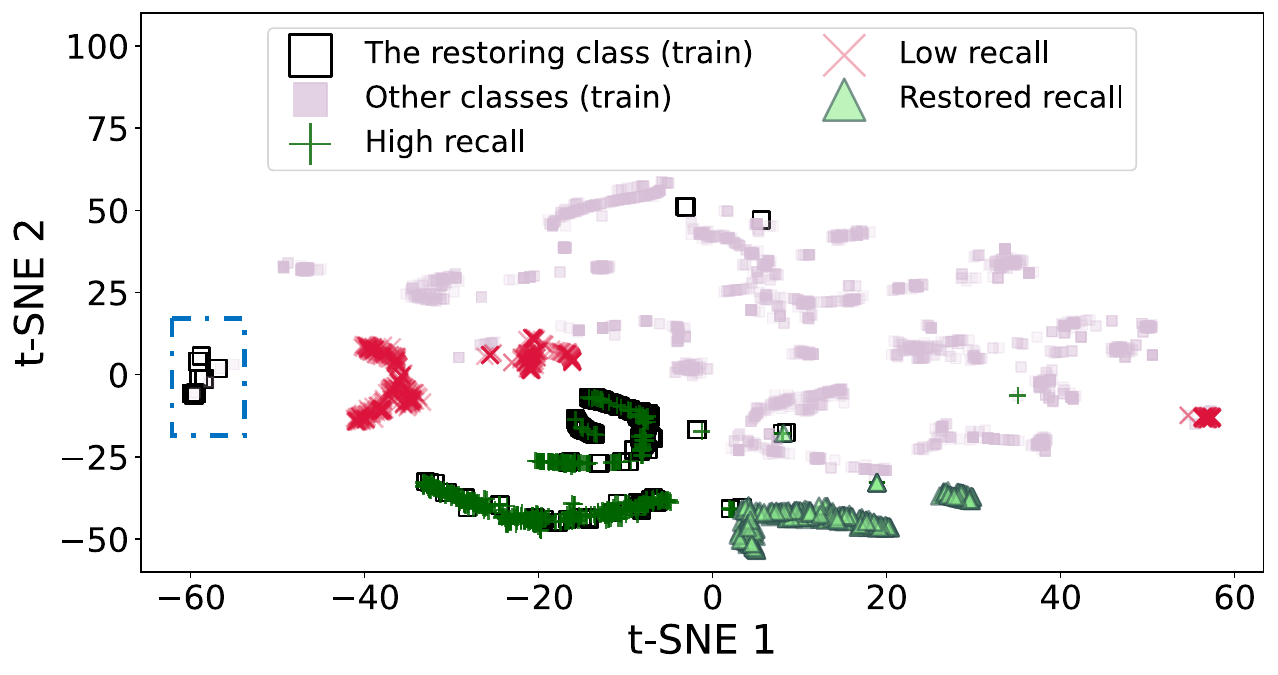}
    \caption{t-SNE visualization for JVC Hub.}
    \label{fig:restoringTSNE}
\end{subfigure}

\caption{Behavioral evolution manifests differently across IoT device classes. Representative examples show persistent behavior (a), behavioral decay (b), and behavioral restoration (c), while the corresponding t-SNE projection (d) illustrates how changes in traffic distribution explain these different evolution patterns. The black curves show the upper-bound maintenance performance obtained by retraining with all newly labeled deployment traffic.}
\label{fig:2x2}
\vspace{-5mm}
\end{figure*}

Behavioral evolution manifests differently across IoT device classes. To characterize this heterogeneity, we analyze the temporal recall of individual device classes. Recall measures the fraction of instances from each class that continue to be correctly identified over time and therefore provides an indirect indication of how closely deployment traffic conforms to the learned behavioral patterns.
Based on the temporal evolution of recall, we observe three representative patterns of behavioral evolution across IoT devices:
\textbf{\textit{(i)}} \textit{persistence} (as exemplified by Philips Hue Bridge in Fig.~\ref{fig:persistentPerf} and demonstrated by five other devices), where no significant drop in recall is observed, \textbf{\textit{(ii)}} \textit{decay} (represented by Google Nest Mini in Fig.~\ref{fig:decayingPerf} and demonstrated by ten other devices), which is characterized by an overall decrease in performance over time, and \textbf{\textit{(iii)}} \textit{restoration} (represented by JVCKENWOOD Hub in Fig.~\ref{fig:restoringPerf} and demonstrated by three other devices), in which the temporal patterns of recall show periods of restoration.

To understand why these different evolutionary patterns arise, we examine the JVCKENWOOD Hub, a representative device that exhibits behavioral restoration. Fig.~\ref{fig:restoringTSNE}, illustrates a t-SNE projection \cite{Hinton:JMLR08} of its traffic across different time periods: 
the training data pertaining to JVCKENWOOD Hub (void boxes), the collective training data of other classes (filled boxes), the test data pertaining to the high recall lasting from 1 Jan'22 to 25 Feb'23 (the plus sign markers), the low performance period of 25 Feb'23 to 5 Aug'23 (the crosses) and the restoration period of 5 Aug'23 to the end of the test phase (triangles). Note that, given the purpose of our analysis, for the two high-performing periods and the low-performing period, we have only included the correctly and incorrectly classified instances, respectively. The training data for other classes is included in deriving the t-SNE to help with contrasting and defining the patterns for JVCKENWOOD Hub against the rest of the classes. 

During the high-performing period, deployment traffic conforms to the behavioral patterns learned during training, even though it is not statistically identical to the training traffic.
For example, no test instances fall into the data mode indicated by the dash-dotted box on the left side of Fig.~\ref{fig:restoringTSNE}. 
In contrast, during low performance periods, instances deviate from the training distribution and partially overlap with other classes (filled boxes).
These observations suggest that the key question is not whether deployment traffic is statistically identical to the training data \cite{Gustavo:KDD16,Qahtan:KDD15}, but whether it continues to \textbf{conform} to the behavioral patterns captured during training. 

During the second high-performance period (triangles), instances exhibit new yet consistent patterns that differ from those observed during training. This shows that behavioral evolution does not necessarily lead to classifier degradation. 
Instead, deployment traffic may evolve while remaining sufficiently consistent with the learned decision boundaries to preserve classification performance. {\color{black}These observations suggest that 
maintaining classifier performance requires more than correcting misclassified instances. As device behavior evolves, newly emerging behavioral patterns should also be incorporated into the learned representation of each device class. The following subsection demonstrates that such maintenance is feasible when freshly labeled deployment traffic is incorporated into classifier retraining}.

\subsection{Classifier Maintenance Using Fresh Traffic}
\label{subsec:adaptationContribution1}
The previous analysis showed that IoT traffic behaviors evolve throughout long-term deployment. A natural question, therefore, arises: how can classifier performance be maintained as these behaviors evolve? 
To answer this question, we first consider an idealized setting in which every deployment-phase traffic instance is labeled and incorporated into model retraining. Specifically, at the end of each 7-day deployment window, all newly observed traffic is labeled and appended to a cumulative training dataset, which is then used to retrain the classifier. Then, the model is assessed on the subsequent window.
Although such exhaustive labeling is operationally expensive, it establishes a practical upper bound on the performance achievable through classifier maintenance.

\begin{figure}[!t]
  \centering
  \includegraphics[width=0.975\linewidth]{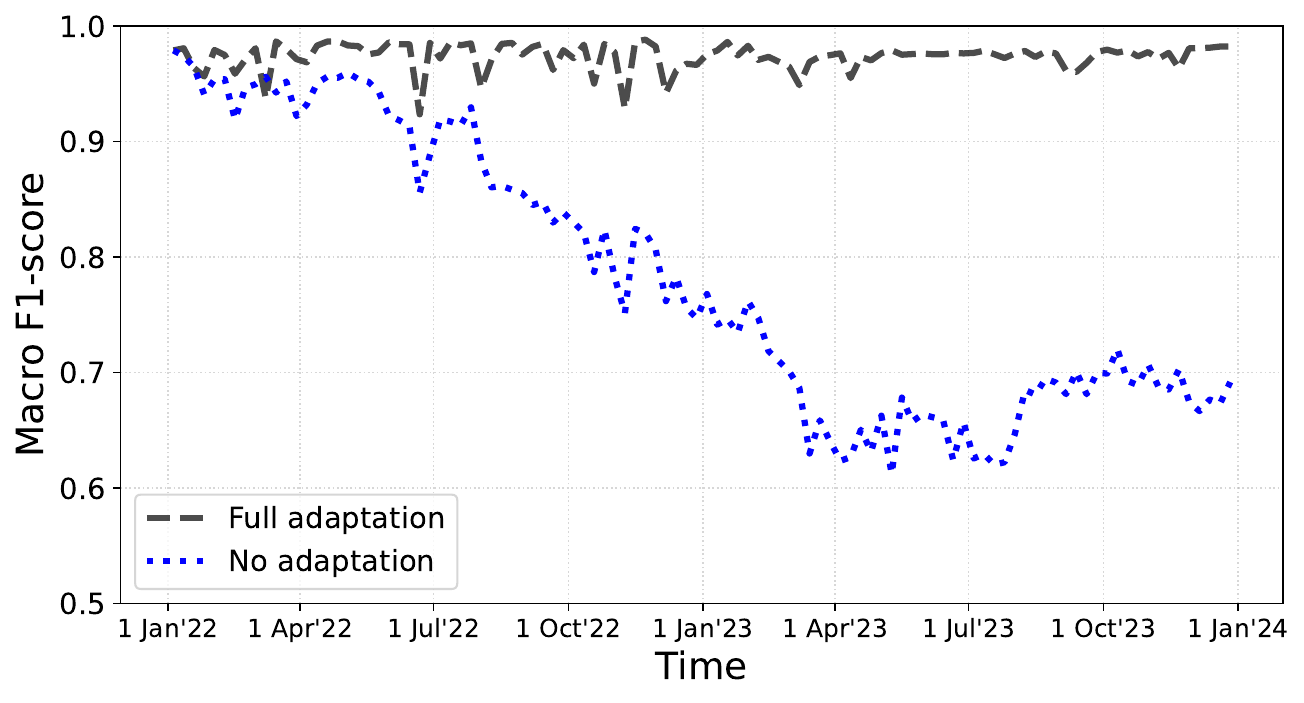} 
  \vspace{-2mm}  
  \caption{Upper-bound maintenance performance established by retraining the classifier using exhaustively labeled deployment traffic. This experiment demonstrates that long-term classifier maintenance is feasible, but highlights the prohibitive labeling effort required in practice.}
  \label{fig:uniformLabeling}
  \vspace{-5mm}
\end{figure}

The resulting upper-bound maintenance performance is shown by the dashed curve in Fig.~\ref{fig:uniformLabeling}.
Retraining the classifier using all newly labeled deployment traffic almost completely eliminates the degradation observed in Fig.~\ref{fig:performanceDrop}.
In particular, the recall profiles of all device classes become persistent after maintenance, indicating that the classifier successfully incorporates the evolving behavioral patterns observed during deployment. 
These results establish that long-term classifier maintenance is feasible by continuously retraining on newly labeled deployment traffic. However, achieving this performance requires labeling approximately 3.4 million traffic instances over the two-year deployment period, corresponding to about 33,000 instances per week on average. 
Such labeling requirements are operationally prohibitive in practice \cite{trainingWith1mille:CCS25}.

The remaining challenge is, therefore, how long-term classifier maintenance can be achieved under realistic labeling constraints. Practical network operators can only label a small fraction of deployment traffic, making exhaustive maintenance infeasible. This raises a practical question: how can evolving behaviors be identified from deployment traffic to enable classifier maintenance without relying on exhaustive labeling?

%% file: sections/5distribution.tex
\section{Conformity-Based Drift Detection in IoT Traffic Features}\label{sec:detection}

\begin{figure*}[!b]
  \centering
  \includegraphics[width=0.90\linewidth]{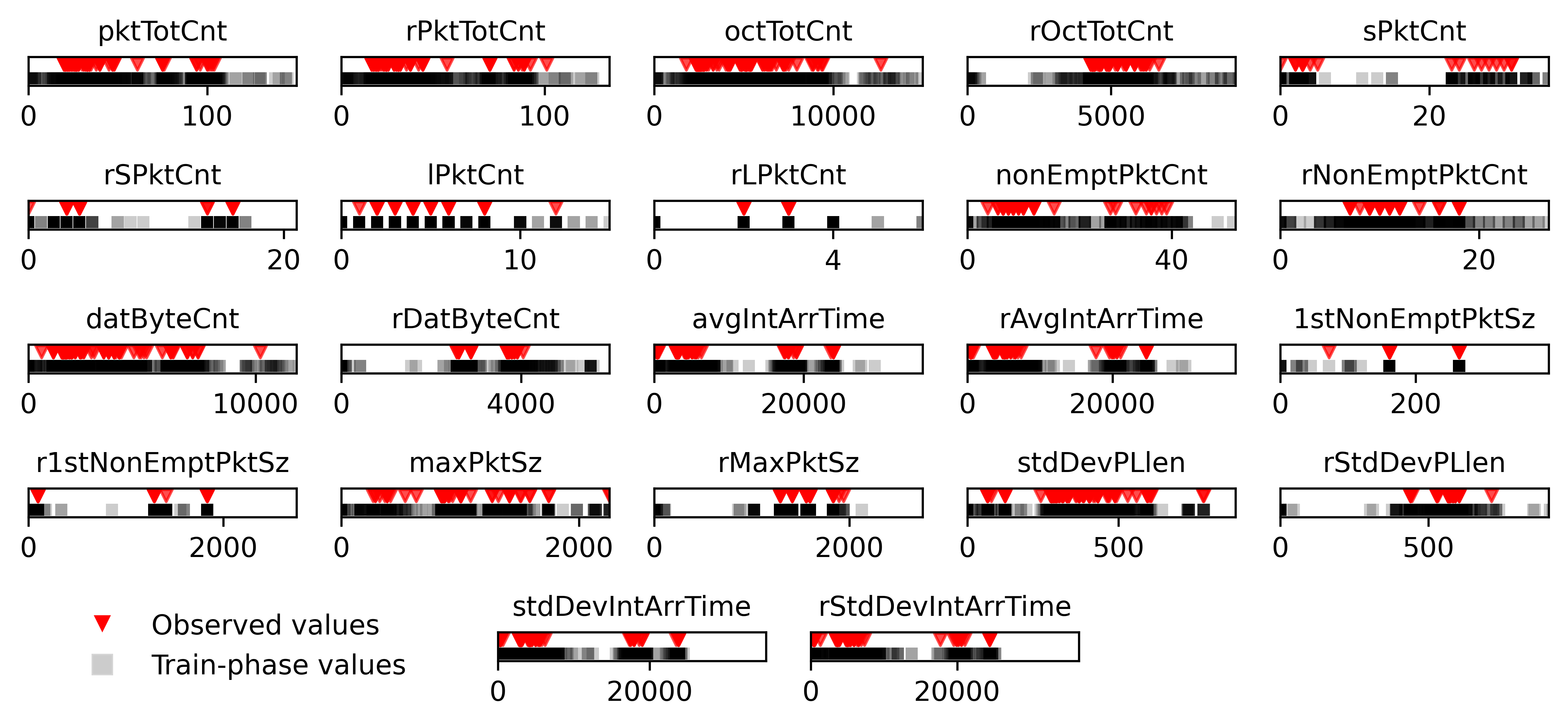} 
  \vspace{-2mm}
  \caption{Feature-level view of a correctly classified deployment window for Philips Hue Bridge. Although the observed feature values (red triangles) occupy only a subset of the training observations (gray markers), they remain within previously learned behavioral patterns, illustrating behavioral conformity despite distributional differences.}
\label{fig:conformityOfCorrectlyClfd}
\vspace{-4mm}
\end{figure*}

The previous section established that behavioral evolution affects deployed classifiers and that classifier maintenance is feasible given sufficient labeled traffic.  The remaining challenge is therefore to efficiently identify behavioral evolution from deployment traffic so that classifier maintenance can be achieved under limited labeling resources. 

To address this challenge, we develop a conformity-based detector that operates directly on raw traffic features. Unlike existing approaches that infer drift from classifier confidence scores or learned feature embeddings, our method models class-conditional behavioral patterns and detects behavioral evolution through feature-level non-conformity. 
The detector is architecturally decoupled from the underlying classifier: it uses the predicted class only to select the corresponding behavioral model, while the detection itself is performed entirely in the original feature space. This design enables feature-level explanations of behavioral evolution and provides the behavioral signal required for practical classifier maintenance.

\subsection{Drift As Seen in Raw Traffic Features}\label{sec:driftsRawFeatures}
The previous section showed that behavioral evolution is reflected through changes in the network traffic generated by IoT devices.
Now, our goal is to characterize behavioral evolution directly from raw traffic features.  
Existing approaches in the data mining literature \cite{Gustavo:KDD16,Qahtan:KDD15,SanjayRanka:KDD07,Lipton:NeurIPS19}, typically quantify changes by measuring distributional divergence between training and deployment data. Although such measures provide a notion of global distributional change, they are less suited to characterizing behavioral evolution in network traffic, which is often heterogeneous and multi-modal. 

To illustrate this limitation, we first consider the Philips Hue Bridge during the first week of deployment, where the classifier maintains nearly perfect recall (Fig.~\ref{fig:persistentPerf}). The corresponding feature distributions are shown in Fig.~\ref{fig:conformityOfCorrectlyClfd}. Although the deployment observations occupy only a subset of feature values seen during training, every observed value remains within previously learned behavioral patterns. In other words, deployment traffic conforms to the training behavior without reproducing the complete training distribution. A distributional divergence measure would nevertheless report a noticeable difference between the training and deployment data because several modes observed during training are absent during the deployment window, even though the classifier maintains nearly perfect recall. These observations suggest that behavioral evolution should be characterized through \textit{conformity} to previously learned behavioral patterns   rather than strict statistical equivalence of traffic distributions.

\begin{figure}[t]
\centering
\subfloat[A Google Nest Mini flow misclassified as an Apple HomePod.]{%
  \includegraphics[width=0.475\textwidth]{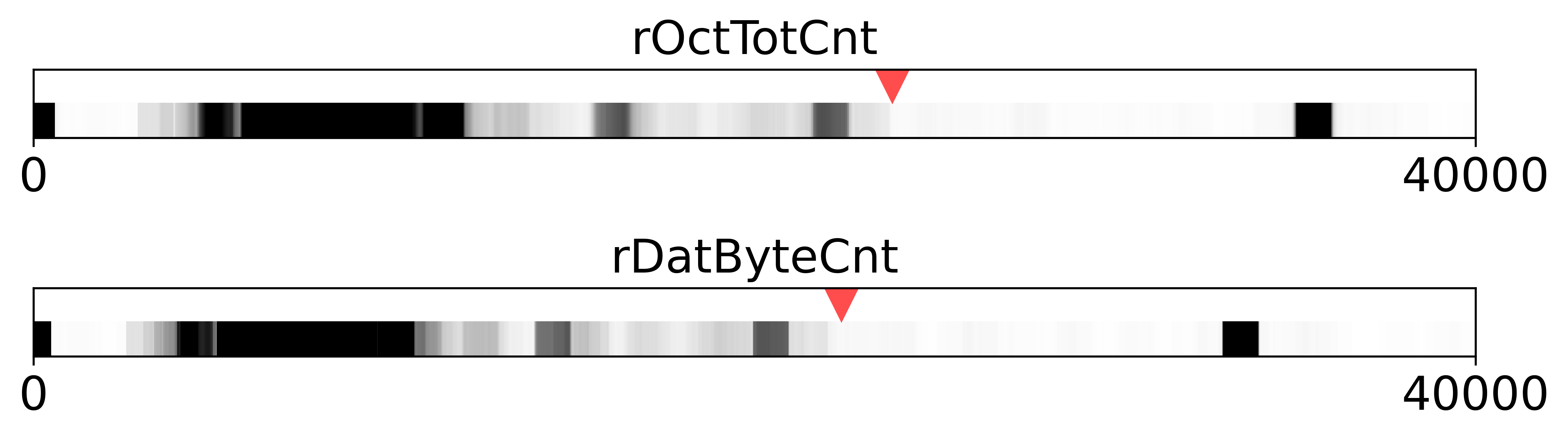}
  \label{fig:nestMiniAsAppleHomepod}
}
\hfill
\subfloat[A misclassified Sony Speaker flow as an Apple HomePod.]{%
  \includegraphics[width=0.45\textwidth]{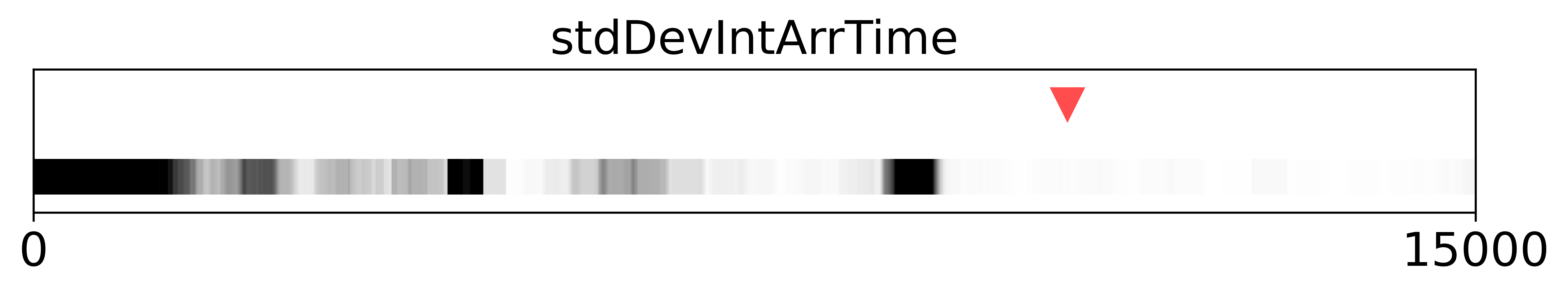}
  \label{fig:sonySpeakerAsAppleHomepod}
}
\caption{Drifted instances reveal feature-level non-conformity with respect to the predicted class. Even when both flows are misclassified as Apple HomePod, different features expose their deviation from the training distribution.}
\label{fig:nonConformitiesInDriftInstances}
\vspace{-5mm}
\end{figure}

Behavioral evolution that leads to classifier degradation exhibits a different characteristic. 
As a representative example, we consider a Google Nest Mini flow collected during the period of behavioral decay (Fig.~\ref{fig:decayingPerf}) that is misclassified as an Apple HomePod. As illustrated in Fig.~\ref{fig:nestMiniAsAppleHomepod}, only two feature values (\ie {\myverbB{rOctTotCnt}} and {\myverbB{rDatByteCnt}}) fall outside the behavioral patterns learned for the predicted class, while the remaining features do conform.
We next consider another deployment instance in which a Sony Speaker flow is misclassified as an Apple HomePod. Unlike the previous example, only a single feature, {\myverbB{stdDevIntArrTime}}, exhibits non-conformity with respect to the learned behavioral patterns (Fig.~\ref{fig:sonySpeakerAsAppleHomepod}).
The key observation here regarding the behavioral evolutions that lead to misclassification is that an instance may move into the decision region of an incorrect class while conforming to that class \textit{only partially}, rather than completely. This partial conformity provides the basis for detecting behavioral evolution from raw traffic features.

The provided examples show that non-conformities are both localized (\ie for a given instance, only a subset of the traffic features typically exhibits non-conformity) and heterogeneous (\ie different instances may evolve through different subsets of features). 
Furthermore, some behavioral changes may not immediately affect classifier performance.
While feature-level conformity should not be viewed as a complete description of behavioral evolution, it provides an interpretable approximation that captures many practically relevant behavioral changes directly in the original feature space. The remainder of this section develops and evaluates a principled and computationally efficient way to identify these non-conforming behaviors.

\subsection{Drift Instance Detection}\label{subsec:detectorFunctionality}
Building on the notion of conformity introduced in the previous subsection, we now develop an instance-level detector to identify behavioral evolution in deployment traffic. The key idea is to compare each feature value of a deployment instance against the behavioral patterns learned for its predicted device class. An instance is regarded as behaviorally non-conforming if one or more of its feature values fall outside the corresponding class-conditional conformity regions. 

To realize this idea, we model the behavioral range of every feature for each device class. Using the most recent labeled traffic, we estimate for every class $c$ and feature $f$ a probability density function (PDF) $\hat{p}_{f,c}$ together with a conformity threshold $\theta_{f,c}$. During deployment, given a feature vector $\textbf{x}=(x_1, x_2,...,x_n)$, we first apply the traffic classifier $M$ to obtain the predicted class $M(\textbf{x})$. The detector then evaluates each feature against the behavioral model for that predicted class. If any feature is non-conforming, the instance is flagged as exhibiting behavioral evolution: 
$D(\textbf{x};M)= 
\bigvee_{i=1}^{n} \mathbf{1}_{\hat{p}_{f_i,M(\textbf{x})}<\theta_{f_i,c}}
$. 
In the following subsections, we explain how the behavioral models are constructed from raw traffic features and how the corresponding conformity thresholds are determined.

\subsection{Modeling Feature Distributions}\label{subsec:modelingFeatDistr}

Having established that behavioral evolution can be characterized by feature-level conformity, we now require a model that represents the behavioral range of each traffic feature. Such a model should faithfully capture the variability observed during training while remaining sufficiently flexible to accommodate the heterogeneous and multi-modal nature of IoT traffic. We, therefore, model the behavior of each feature independently using probability density functions estimated directly from the training data.
Methods for modeling data distributions are broadly categorized as parametric and non-parametric. Parametric methods assume a predefined family of distributions, which is unsuitable for complex and multimodal traffic features. Non-parametric methods, in contrast, allow the data itself to define the shape of the distribution and are therefore better suited to our setting. 
Among non-parametric approaches, we require a method that captures complex feature distributions while remaining computationally efficient and interpretable.
We adopt Kernel Density Estimation (KDE), where the density of a random variable $X$ given observations $x_1, x_2, ..., x_m$ is estimated as: $\hat{p}_X(x) = \frac{1}{mh}\sum_{i = 1}^{m}k\left(\frac{x-x_i}{h}\right)$, where $k(\cdot)$ is a kernel function and $h$ is the bandwidth controlling smoothness. 
Applying KDE to IoT traffic features requires addressing two practical challenges: (1) choosing a valid kernel function and (2) handling zero-valued and degenerate observations.

\begin{figure*}[!t]
    \centering
    \captionsetup{skip=1mm}
    \subfloat[GKDE.\label{fig:GKDE}]{
        \includegraphics[width=0.95\columnwidth]{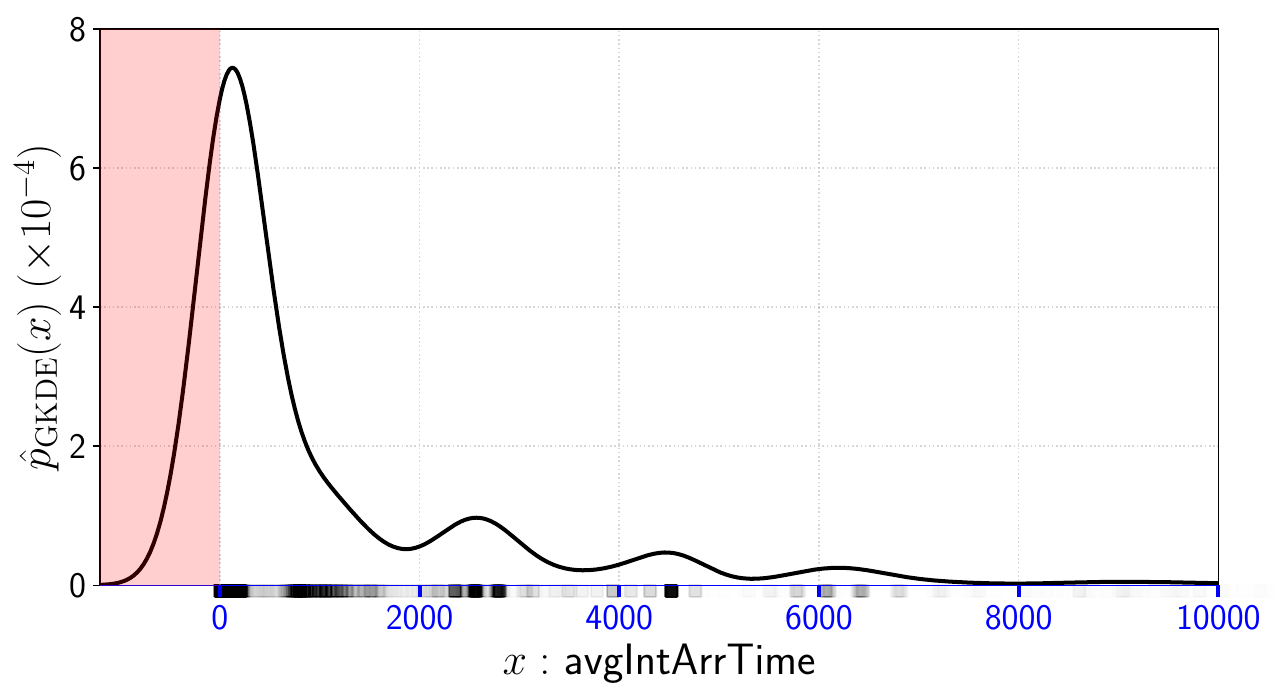}
    }
    \hfill
    \subfloat[LNKDE.\label{fig:LNKDE}]{
        \includegraphics[width=0.95\columnwidth]{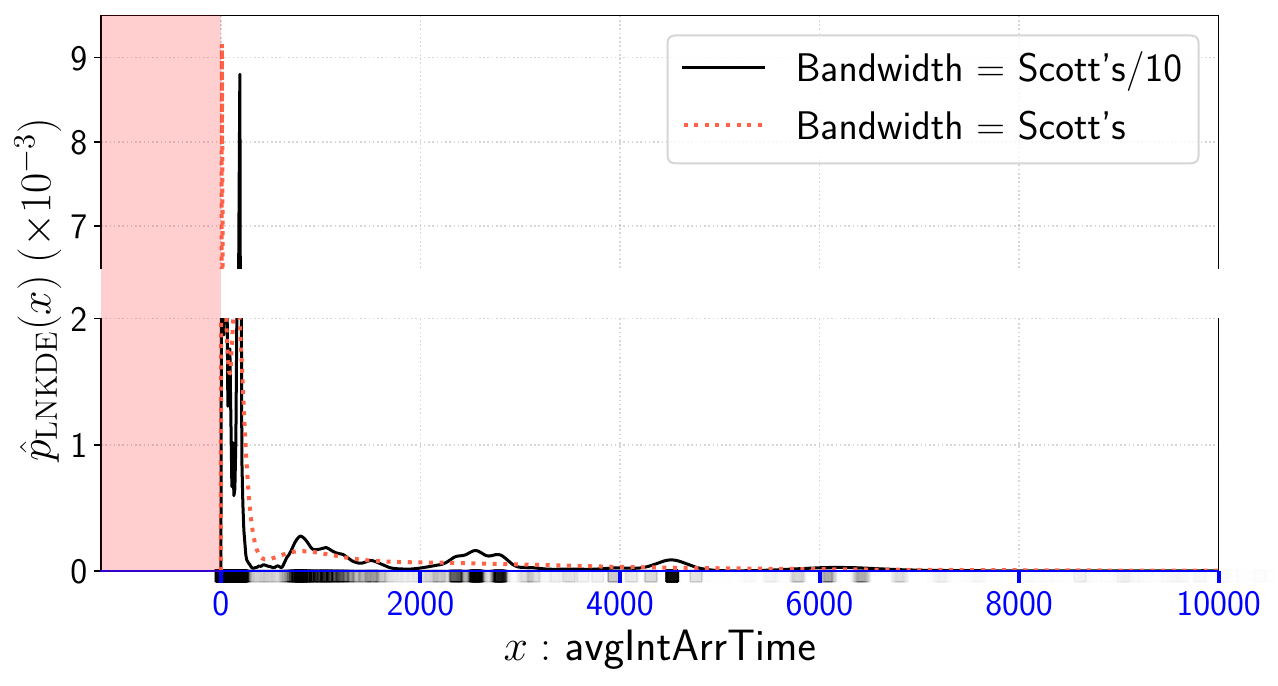}
    }
    \caption{Modeling the behavioral range of non-negative traffic features. (a) GKDE assigns non-zero probability to infeasible feature values ($x<0$), which is invalid for non-negative features. (b) Log-normal KDE (LNKDE) preserves non-negativity while accurately modeling the observed behavioral range of the feature.}
    \label{fig:KDE}
\end{figure*}

\textbf{The choice of kernel $k$.} A common choice of $k$ is the Gaussian function, where $k(t) = \frac{1}{\sqrt{2\pi}} e^{-\frac{t^2}{2}}$, and the resulting method is called Gaussian Kernel Density Estimation (GKDE). A behavioral model should assign probability only to feasible traffic values. However, Gaussian kernels are symmetric and assign non-zero density to $x<0$, which is invalid for non-negative traffic features. To illustrate this limitation,Fig.~\ref{fig:GKDE} shows the density estimated for the {\myverbB{averageInterarrivalTime}} feature of Apple Homepod flows using GKDE. The bandwidth $h$ is chosen using Scott's rule-of-thumb \cite{scott}. The mass of this invalid region, as estimated by the resulting distribution, is $0.24$. 

We address this by applying a log transformation of $Y = \log X$ \cite{charpentier:halshs-01115988,Nguyen:logKDE}, estimating the density in log-space, and transforming it back according to $\hat{p}_X(x) = (\log x)^\prime \hat{p}_Y(\log x) = \frac{\hat{p}_Y(\log x)}{x}$. This yields a log-normal KDE (LNKDE) that preserves the non-negative support of the traffic feature while retaining its behavioral interpretation.
Fig.~\ref{fig:LNKDE} illustrates the LNKDE corresponding to Fig.~\ref{fig:GKDE}, which results in a density of zero for values of $X\leq0$. For the bandwidth $h$ when constructing $\hat{p}_Y(y)$, we scale Scott's rule-of-thumb by a factor of $0.10$ (solid curve) to avoid over-smoothing (dotted curve). In this paper, we use Scott's rule of thumb divided by $10$, since it yields quality estimates in practice. In both Fig.~\ref{fig:GKDE} and Fig.~\ref{fig:LNKDE}, the scatter plot of observed samples is provided as a reference on the x-axis.

\textbf{Zero values and degeneracies.} The $\log$-transformation is not applicable to features that can take the value $X=0$. To handle this, we model the distribution as a combination of a degenerate component (point mass) at $X=0$ and a continuous component for non-zero values. The point mass captures zero-valued observations, while the non-zero component is modeled using LNKDE when sufficient variability is present. 

Similarly, when the non-zero observations are singular, we use a degenerate distribution at the observed value since Scott's bandwidth relies on the variance of observations and is therefore not applicable. This leads to a total of 5 modeling scenarios depending on whether zero values and/or singularities are present, as listed in Appendix~\ref{app:distributionModelingCases}. The vast majority of the distributions are modeled either solely by an LNKDE component or by an LNKDE component together with a degeneracy at $X=0$.

\subsection{Constructing Feature Conformity Regions}
\label{subsec:conformityThreshold}
Having constructed a behavioral model for each traffic feature, the next step is to determine whether an observed feature value continues to conform to the learned behavior. We therefore seek a conformity region for every class-feature pair that separates behavior consistent with the training data from behavior that is sufficiently uncommon to indicate behavioral evolution.
Given the estimated density $\hat{p}_{f,c}$, we define the non-conforming region as the set of feature values with low probability density:  
$\chi_{\theta_{f,c}} = \{x|\hat{p}_{f,c}(x)\leq \theta_{f,c}\}$ 
and the conforming region as its complement: $\chi_{\theta_{f,c}}'=\{x|\hat{p}_{f,c}(x)> \theta_{f,c}\}$.

Because the behavioral ranges of different traffic features vary substantially across device classes, a universal notion of ``low-density'' is not meaningful, making it difficult to choose a universal threshold $\theta_{f,c}$. At the same time, manually specifying thresholds for each feature and class is impractical, especially where distribution models are continuously updated. We therefore automatically derive $\theta_{f,c}$ from the shape of the corresponding distribution.

To determine an appropriate threshold, we constrain the probability mass of the non-conformity region. 
Specifically, we require that the probability of observing values in $\chi_{\theta_{f,c}}$ does not exceed a predefined value $\epsilon$, \ie 
$\int_{\chi_{\theta_{f,c}}}\hat{p}_{f,c}(x')dx' \leq \epsilon$. 
The parameter $\epsilon$ controls the probability mass in the non-conforming region and thus the detector's sensitivity. 

The trivial solution $\theta_{f,c}=0$ satisfies this constraint but results in all observations being classified as conforming, thereby failing to detect drift.
Therefore, we seek the largest value of $\theta_{f,c}$ that satisfies the constraint, leading to the following optimization problem:
\begin{equation}
\label{eq:optimization}
\begin{aligned}
\text{maximize} \quad & \theta_{f,c} \\
\text{subject to} \quad & \int_{\{x|\hat{p}_{f,c}(x)\leq\theta_{f,c}\}}\hat{p}_{f,c}(x')dx' \leq \epsilon.
\end{aligned}
\end{equation}
\noindent Note that the integral above also gives the probability of falsely flagging an instance drawn from $\hat{p}_{f,c}(x)$. The parameter $\epsilon$ is the only hyperparameter in our methodology that needs to be set by the user.

\begin{figure}[!t]
 \centering
 \includegraphics[width=0.95\linewidth]{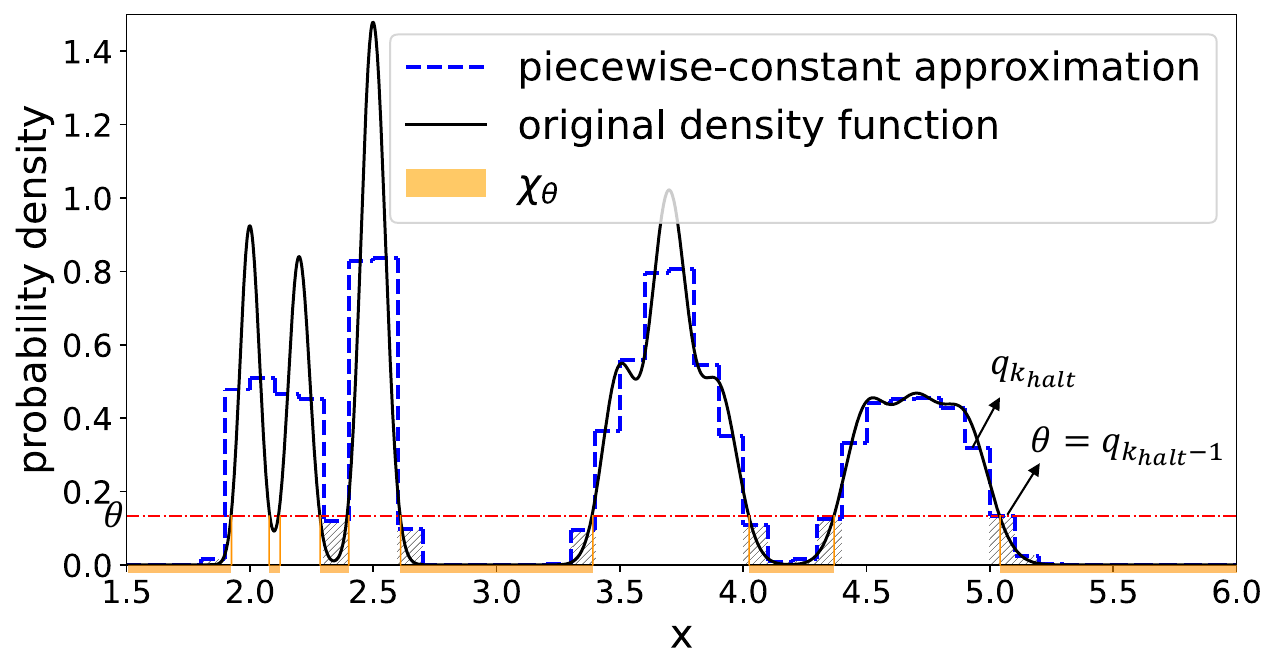}
  \vspace{-3mm}
  \caption{Construction of feature-level conformity regions. The conformity threshold $\theta$ is chosen so that the probability mass of the non-conforming region (shaded) does not exceed the user-defined sensitivity parameter $\epsilon$.}
\label{fig:theta_epsilon_methodology}
\vspace{-4mm}
\end{figure}

To compute the conformity threshold $\theta_{f,c}$, we use a piecewise-constant approximation of the density function. 
We discretize the domain into bins $\mathcal{B}=\{(b_{1},b_{1}+\Delta), (b_{2},b_{2} + \Delta), ..., (b_{B},b_{B} + \Delta)\}$ 
estimate the density in each bin as $p_i = \frac{1}{\Delta} \int_{b_i}^{b_i+\Delta} \hat{p}_{f,c}(x) \, dx.$
We then sort these values in ascending order and accumulate the probability mass until the constraint $\epsilon$ is exceeded. The threshold $\theta_{f,c}$ is selected as the largest density value that satisfies the constraint. 
The detailed procedure is provided in Algorithm~\hyperlink{alg:findtheta}{1}.

Fig.~\ref{fig:theta_epsilon_methodology} illustrates this approximation for $\epsilon = 0.1$. The piecewise constant approximation (blue dashed lines) is shown alongside the original density (the solid black curve), and the resulting $\theta_{f,c}$ separates conforming and non-conforming regions. 

Finally, we note that the above formulation assumes the densities are modeled using LNKDE.  
For degenerate distributions,  
the threshold is set to $0$, ensuring that only values at the support points are considered conforming. 
In mixed cases, degenerate components are treated as conforming at their support, while the thresholding procedure is applied to the continuous component.

{\color{black}
\subsection{Behavioral Non-conformity and Classifier Misclassification}
We next investigate the relationship between behavioral non-conformity and classifier misclassifications.
To quantify this relationship, we evaluate the detector using the True Positive Rate (TPR), defined as the fraction of misclassified instances that are detected, and the False Positive Rate (FPR), defined as the fraction of correctly classified instances that are incorrectly flagged as exhibiting behavioral evolution. We perform the evaluation across the entire deployment phase, fitting $M(\textbf{x})$ and $D(\textbf{x}; M)$ to the training data and keeping both unchanged throughout the evaluation. We start by reporting the results for the selected sensitivity threshold of $\epsilon=0.01$. A discussion on the role of $\epsilon$ is provided in the following subsection.
}

\begin{algorithm}[t!]
\hypertarget{alg:findtheta}{}
\caption{Conformity Threshold Calculation}
\begin{algorithmic}
\STATE \textbf{Input:} PDF $\hat{p}(x)$, constant $\epsilon$ and a binning $\mathcal{B}=\{(b_{1},b_{1}+\Delta), (b_{2},b_{2} + \Delta), ..., (b_{B},b_{B} + \Delta)\}$, where $b_i=b_{i-1}+\Delta$.
\STATE \textbf{Output:} Threshold $\theta$.
\FOR{$i = 1$ to $B$}
\STATE $p_i \leftarrow \frac{\int_{b_{i}}^{b_{i}+\Delta} \hat{p}(x) dx}{\Delta}$
\ENDFOR
\STATE $q_1,q_2,...,q_B \leftarrow \text{sort}\_\text{ascending}(p_1,p_2,...,p_B)$
\STATE $\theta_{prev}=0$
\FOR{$k = 1$ to $B$}
\STATE $m\leftarrow \Delta\sum_{k'=1}^{k}q_{k'}$

\IF{$m>\epsilon$}
\STATE $\theta \leftarrow \theta_{prev}$
\STATE \textbf{break}
\ELSE
\STATE $\theta_{prev} \leftarrow q_k$
\ENDIF
\ENDFOR
\STATE \textbf{return} $\theta$ 
\end{algorithmic}
\end{algorithm}

\begin{figure*}[!t]
    \centering
    \subfloat[True Positive Rate.\label{fig:detectionTPR}]{
        \includegraphics[width=0.95\columnwidth]{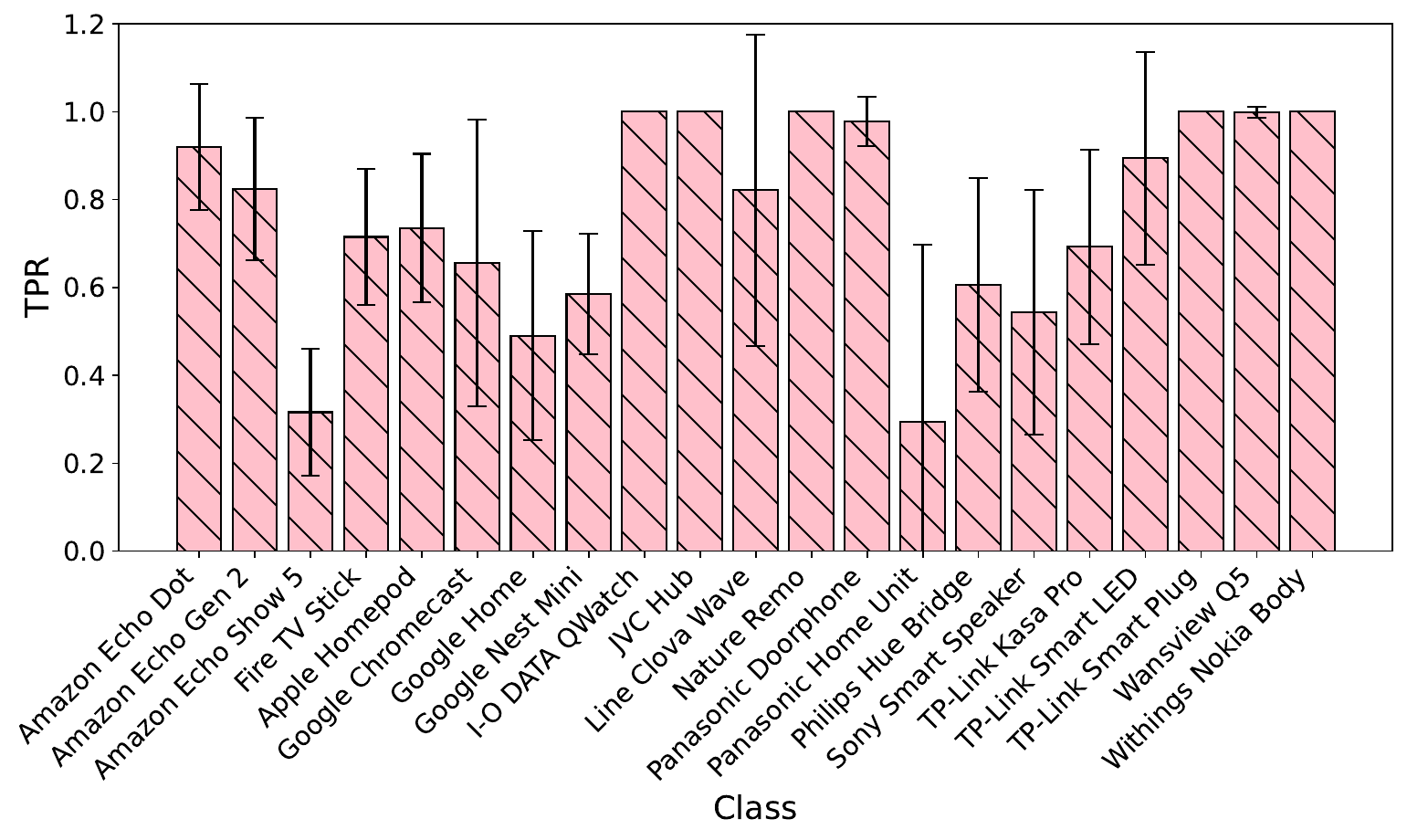}
    }
    \hfill
    \subfloat[1 - False Positive Rate.\label{fig:detectionSpecificity}]{
        \includegraphics[width=0.95\columnwidth]{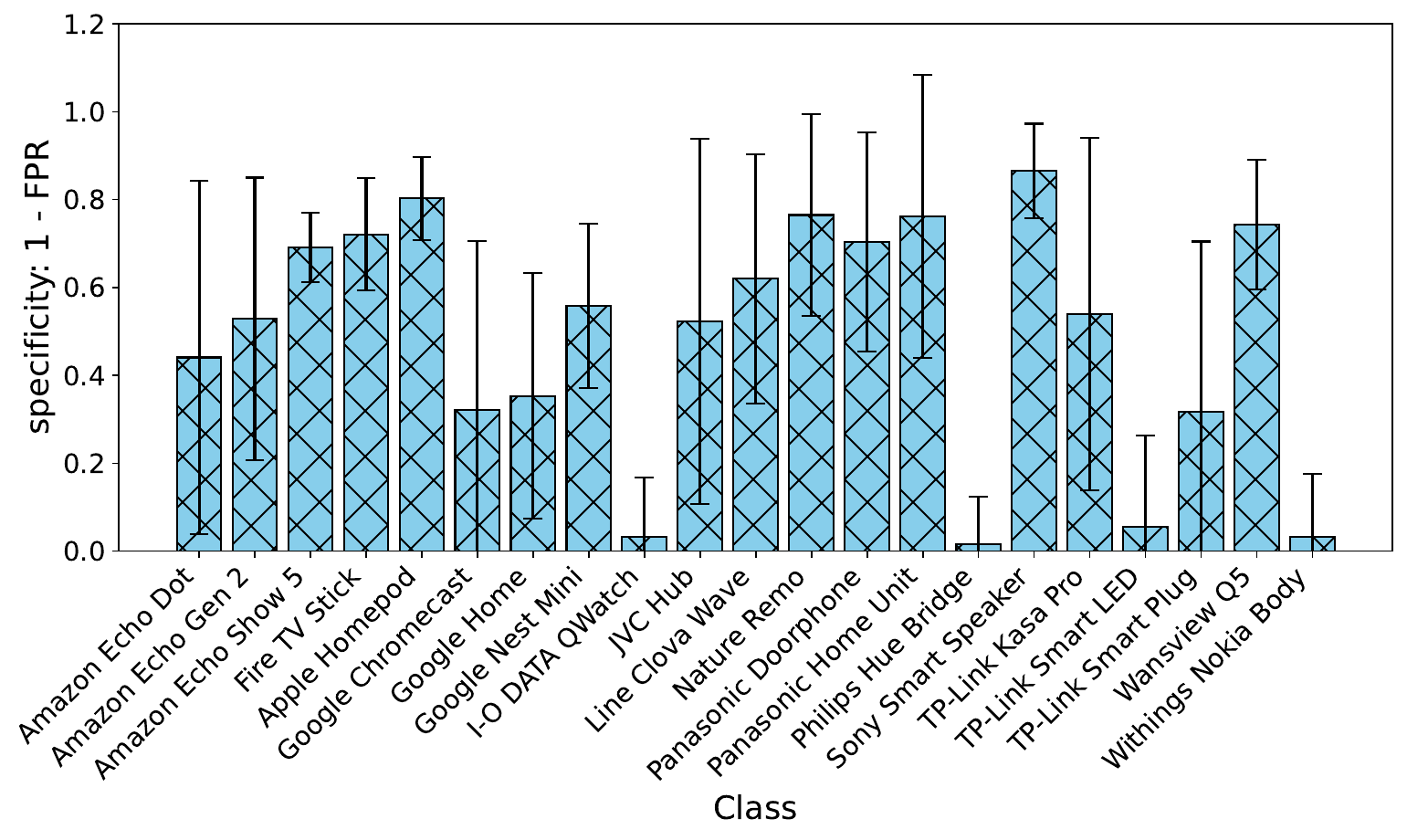}
    }
    \vspace{-1mm}
    \caption{Relationship between behavioral non-conformity and classifier misclassification. Mean TPR (a) and specificity (b) are reported for each device class over the deployment phase. Error bars indicate temporal standard deviation across evaluation windows.}
    \label{fig:detectorTPR_FPR}
    \vspace{-6mm}
\end{figure*}

{\color{black}
Fig.~\ref{fig:detectorTPR_FPR} summarizes the detector performance across device classes, averaged over the deployment period. We report mean TPR and specificity ($1-\mathrm{FPR}$) averaged over weekly evaluation windows, with error bars indicating the temporal standard deviation. As shown in Fig.~\ref{fig:detectionTPR}, the detector captures a substantial fraction of the misclassified traffic for most device classes, although the achieved TPR varies across classes and over time.  
For several devices (\eg Line Clova Wave and Panasonic Home Unit), the TPR exhibits noticeable temporal variation. This indicates that behavioral evolution manifests differently over time across device classes. Averaged across all device classes, the detector achieves a temporal mean TPR $0.77$.} 

\begin{figure}[!t]
 \centering
 \includegraphics[width=0.95\linewidth]{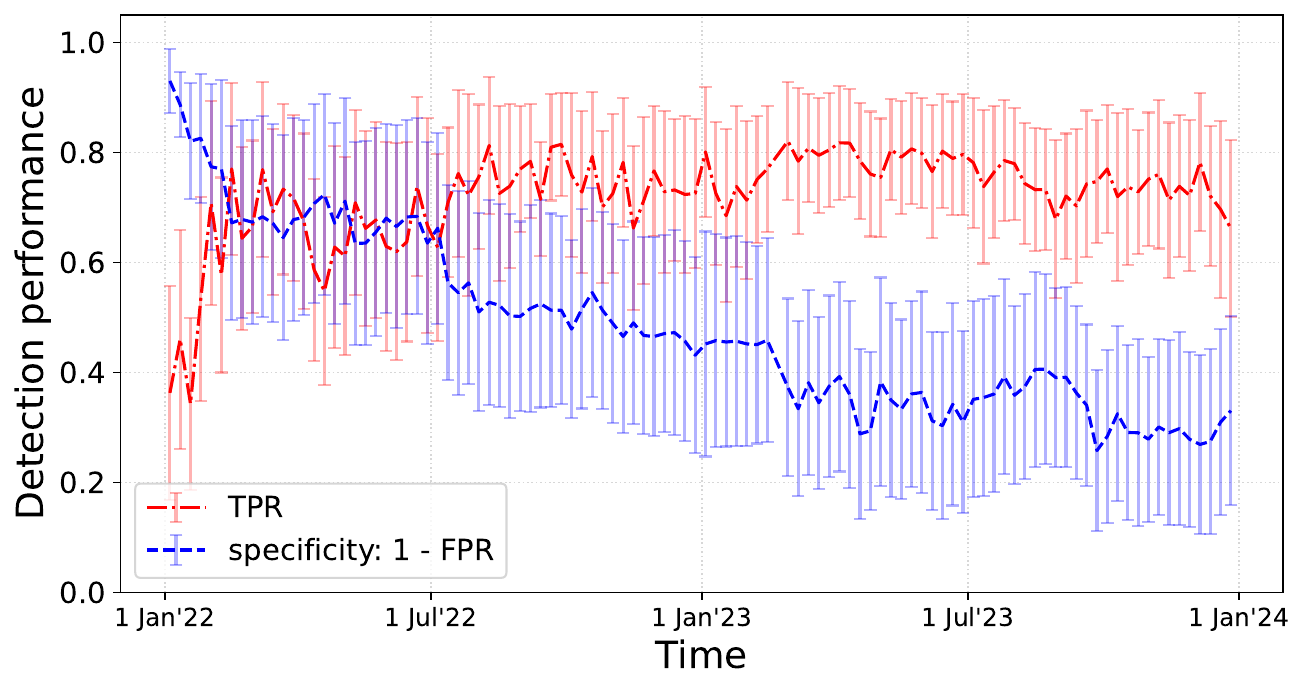} 
  \vspace{-3mm}
  \caption{Temporal evolution of detector performance averaged across device classes. {\color{black}TPR remains relatively stable throughout deployment, whereas specificity gradually decreases as behavioral evolution introduces an increasing number of correctly classified instances exhibiting previously unseen patterns.}} 
\label{fig:TPR_FPR_temporal_trend}
\end{figure}

{\color{black}
Depicted in Fig.~\ref{fig:detectionSpecificity}, specificity results show that when aggregated over the full deployment phase, the detector is prone to flagging correctly classified instances. However, the large error bars suggest that the specificity is highly variable over time. The temporal behavior is shown in Fig.~\ref{fig:TPR_FPR_temporal_trend}. 
Specificity gradually decreases throughout deployment. 
During the early stages of deployment, most correctly classified instances remain behaviorally conforming (the average specificity over the first six months of deployment is $0.71$, with 13 device classes exceeding $0.8$). However, as device behaviors continue to evolve, an increasing number of correctly classified instances exhibit previously unseen behavioral patterns. These emerging patterns do not immediately cause misclassifications, but they are identified by the detector as behavioral non-conformity. Consequently, the detector should be updated periodically to incorporate these newly observed behaviors. 
The relatively stable TPR indicates that a consistent fraction of misclassifications remains detectable based on marginal feature non-conformity throughout deployment. We later show in \S\ref{sec:adaptEval} how this detection behavior affects detector-guided adaptation and how it can be addressed. 
}

\vspace{-0.5mm}
{\color{black}
\subsection{Choice of $\epsilon$}
We evaluate three values of $\epsilon$ spanning three orders of magnitude, which is the only hyper-parameter of our method: $10^{-3}, 10^{-2}$ and $10^{-1}$. The TPR and specificity results, as well as balanced accuracy (defined as $\frac{\mathrm{TPR+Specificity}}{2}$), are shown in Table~\ref{tab:choiceOfEpsilon}. By increasing $\epsilon$, the detection rate increases, though at the cost of more FPs (especially pronounced for $\epsilon=0.1$). In terms of balanced accuracy, different choices meet at a similar point, while $\epsilon=0.01$ slightly outperforms. A more detailed sensitivity analysis around $\epsilon=0.01$ is provided in Appendix~\ref{app:sensitivityEpsilon}.}

\begin{table}[t]
\caption{Detection results for different $\epsilon$ values.}
\label{tab:choiceOfEpsilon}
\centering
\renewcommand{\arraystretch}{1.1}
\begin{tabular}{|c|c|c|c|}
\hline
\textbf{$\epsilon$} & \textbf{TPR} & \textbf{Specificity} & \textbf{Balanced accuracy} \\
\hline
$0.001$ & $0.65\pm 0.27$ & $0.58\pm 0.31$ & $0.61\pm 0.17$ \\
$0.010$ & $0.77\pm 0.22$ & $0.49\pm 0.27$ & $0.63\pm 0.15$ \\
$0.100$ & $0.95\pm0.08$ & $0.26\pm0.19$ &$0.60\pm0.10$ \\
\hline
\end{tabular}
\end{table}

\vspace{-1mm}
\subsection{Feature-Level Explanations of Behavioral Evolution}

Operating directly on raw traffic features naturally provides feature-level explanations for every detected instance. Specifically, the subset of non-conforming features identifies which aspects of the observed traffic behavior deviate from the learned behavioral representation of the predicted device class. Such explanations provide operators with interpretable insights into how device behavior evolves over time.

Each detected instance is associated with a binary indicator vector that identifies the non-conforming features.
Beyond explaining individual detections, these indicator vectors can also be analyzed statistically to assess the consistency of the explanations throughout deployment.  Those analyses are presented in Appendix~\ref{app:explainMore}; here we focus on representative examples. The following example illustrates how these explanations evolve as device behavior changes over time.

\begin{figure}[!t]
\centering
\subfloat[When predicted as Apple HomePod.]{
    \includegraphics[width=0.45\columnwidth]{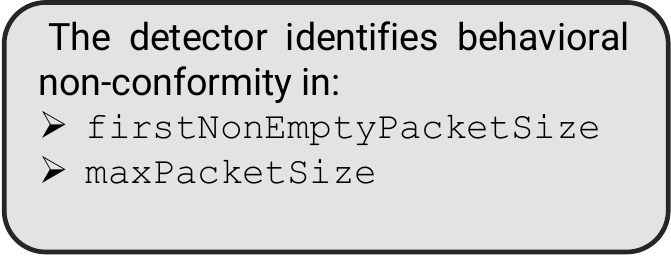}
    \label{fig:exp_as_appleHomepod}
}
\hfill
\subfloat[When predicted as Amazon Echo Dot.]{
    \includegraphics[width=0.45\columnwidth]{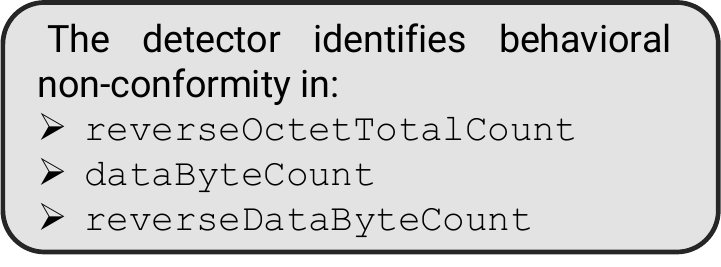}
    \label{fig:exp_as_amazonEchoDot}
}
\caption{Feature-level explanations of behavioral evolution for the Wansview Q5 IP camera. Misclassifications across two deployment periods are explained by distinct subsets of non-conforming traffic features.}
\label{fig:explanations}
\vspace{-3mm}
\end{figure}

To illustrate the detector's explanations, we consider the Wansview Q5 IP camera. Beginning in late 2022, the classifier starts to misclassify this device, initially as an Apple HomePod and later as an Amazon Echo Dot. 
The detector provides distinct feature-level explanations for these two episodes of behavioral evolution, as illustrated in Fig.~\ref{fig:explanations}.

Among the instances misclassified as Apple HomePod, the detector consistently identifies non-conformity 
in the features {\myverbB{maxPacketSize}} and {\myverbB{firstNonEmptyPacketSize}}, as summarized in Fig.~\ref{fig:exp_as_appleHomepod}.
Later in deployment, the same device begins to be misclassified not only as an Apple HomePod but also as an Amazon Echo Dot. These instances are characterized by a different explanation involving {\myverbB{reverseOctetTotalCount}}, {\myverbB{dataByteCount}}, and {\myverbB{reverseDataByteCount}}, as illustrated in Fig.~\ref{fig:exp_as_amazonEchoDot}. This demonstrates that the same device may undergo different forms of behavioral evolution over time.

These examples demonstrate that behavioral evolution is dynamic and instance-specific. Different stages of deployment may be characterized by different subsets of non-conforming traffic features, allowing the detector to provide interpretable feature-level explanations directly in the original feature space.

%% file: sections/6evaluation.tex
\section{Adapting IoT Traffic Classifiers to \\Concept Drift}\label{sec:adaptEval}

The previous sections established that classifier maintenance throughout deployment is necessary and feasible, and introduced a behavioral signal that identifies evolving traffic patterns. The remaining challenge is determining how this behavioral signal should be exploited for long-term classifier maintenance in operational deployments, where only a limited fraction of deployment traffic can realistically be labeled.  
The following subsections investigate this question through a series of adaptation experiments.

\begin{figure*}[!t]
    \centering
    \subfloat[Adaptation performance.\label{fig:classificationPerformance}]{
        \includegraphics[width=0.95\columnwidth]{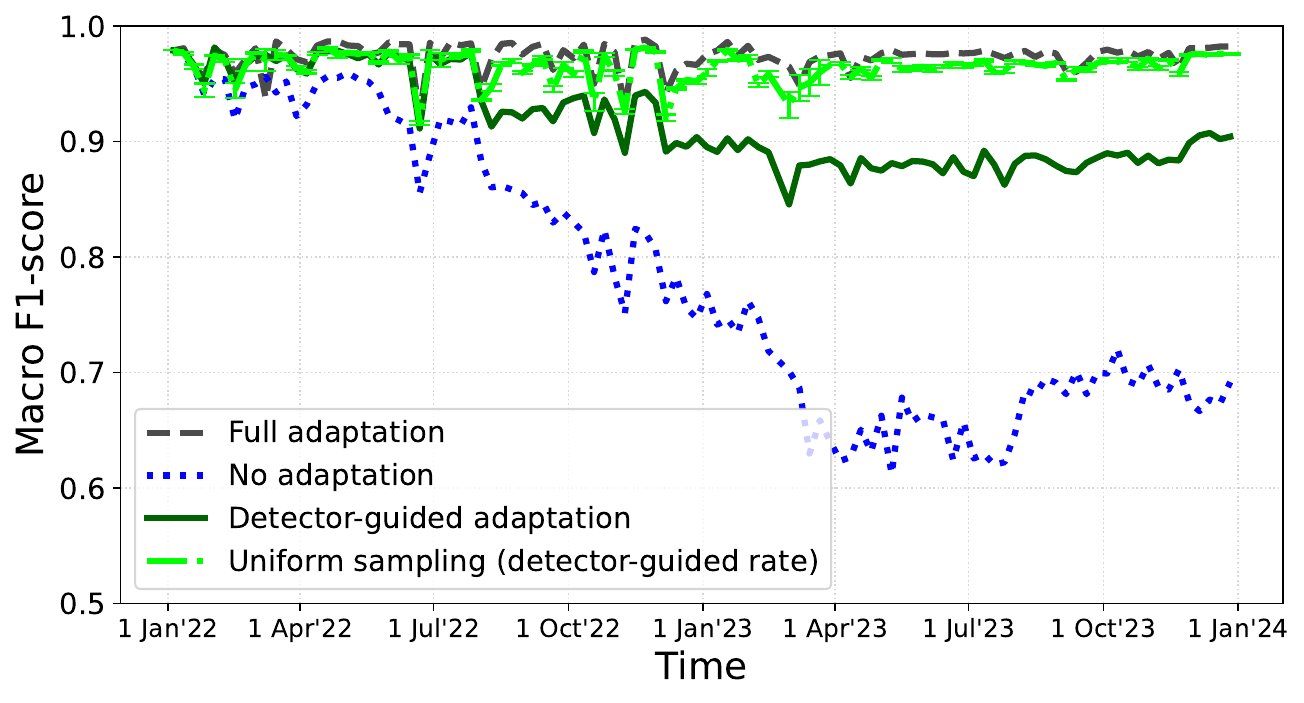}
    }
    \hfill
    \subfloat[Labeling cost.\label{fig:labeling cost}]{
        \includegraphics[width=0.95\columnwidth]{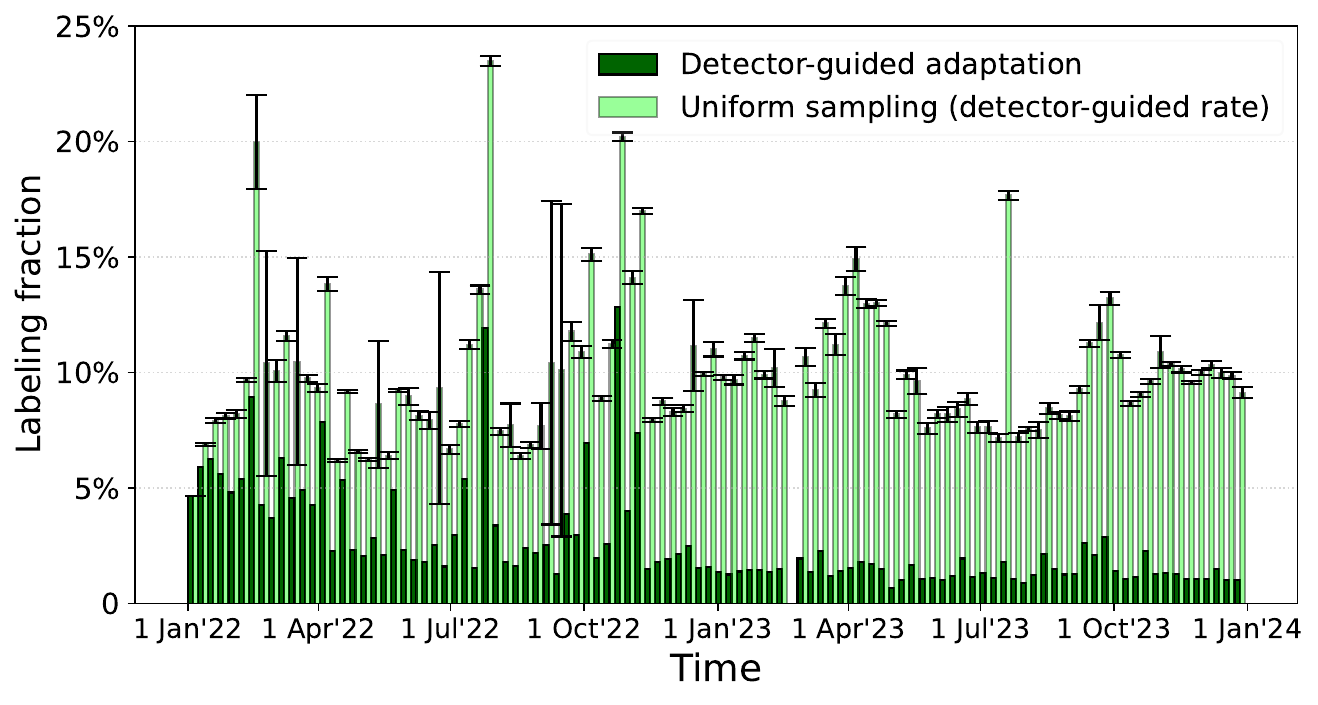}
    }
    \vspace{-1mm}
    \caption{
    {(a) Classification performance under different adaptation strategies. Detector-guided adaptation (solid green) substantially improves performance over no adaptation (dotted blue), but remains consistently below the upper-bound maintenance strategy (dashed gray). Replacing detector-selected retraining instances with uniformly sampled deployment traffic at the same detector-guided labeling rate (dash-dotted green) yields performance close to the upper bound.  
   (b) Fraction of deployment traffic labeled in each evaluation window. Detector-guided adaptation labels progressively fewer instances over time as fewer deployment flows are identified as behaviorally non-conforming, whereas uniform sampling preserves the detector-guided sampling rate while providing more representative coverage of the evolving behavioral patterns. {\color{black}For the strategy involving random sampling, error bars indicate the standard deviation across 10 random seeds.}}
   }
   \label{fig:adaptation_perf_and_labeling}
    \vspace{-4mm}
\end{figure*}

\subsection{Detector-Guided Adaptation}
\label{subsec:detector-guidedAdapt}
We begin by considering the most natural adaptation strategy, in which the deployment instances identified by the detector are labeled and incorporated into classifier retraining. 
Let $W_i$ (for $i>0$) denote the (unlabeled) dataset corresponding to test window $i$, with $i=0$ reserved for training data. We use $W_i^*$ to denote the labeled version of $W_i$  
 (\eg $W^*_{0}$ denotes the labeled training dataset) and define the cumulative dataset up to window $i$ as $W^*_{cum,i}$, 
 initialized as $W^*_{cum,0}=W^*_{0}$. 
 
Let $\mathcal{M}(W^*)$ and $\mathcal{D}(W^*;\epsilon)$ denote procedures for training the classifier $M$ and the drift detector $D$, respectively. The detector construction follows \S\ref{sec:detection}, where feature distributions and conformity thresholds are estimated using parameter $\epsilon$. {\color{black}Throughout this section, we use $\epsilon=0.01$.} 
The classifier is implemented as a Random Forest, with hyperparameters fixed from the training phase (see \S\ref{subsec:classifier}).

We define the drift instance selection function as $\tilde{D}(W;M)=\{\textbf{x}\in W|D(\textbf{x};M)=1\}$. 
Let $M_{i-1}$ and $D_{i-1}$ denote the classifier and the detector after processing test window $i-1$. For window $i$, we first evaluate the performance of $M_{i-1}$ on $W_{i}$. Then we label $\tilde{D}_{i-1}(W_{i};M_{i-1})$ and concatenate the result to $W^*_{cum,i-1}$. Finally, we update both the detector and the classifier using the newly extended cumulative dataset. This procedure is summarized in the Appendix~\ref{app:detector-guidedAdapt}. The resulting adaptation performance is depicted in Fig.~\ref{fig:classificationPerformance}, in terms of macro F1-score. The average of the macro F1-score across all deployment windows is $0.917$. This level of adaptation performance is achieved by labeling $2.66\%$ of the deployment instances overall. Per-window labeling cost is depicted in Fig.~\ref{fig:labeling cost}. 

Although detector-guided adaptation substantially improves classification performance, it does not achieve the level of recovery observed under the upper-bound maintenance strategy. This bias is attributed to the limitation of the detector in identifying all possible pattern changes (recall the TPR results from Fig.~\ref{fig:detectionTPR}), and is particularly manifested in a performance decline from early August 2022 (see Fig.~\ref{fig:classificationPerformance}). To illustrate this bias, we have chosen one of the classes that experienced a performance drop during the mentioned period, namely the Google Nest Mini. We consider a slice of the deployment phase commencing at the performance drop: in the window starting from 30 July 2022, the F1-score for this class drops to $0.446$ from $0.938$ in the preceding window, while the poor performance ($0.389$ F1-score) continues in the subsequent window. 
This indicates the emergence of a novel pattern from the end of July that  is not adequately captured by the detector, since otherwise the degradation would not persist into the subsequent window. 

These observations reveal an important limitation of detector-guided adaptation. Although the detector successfully identifies a substantial fraction of behavioral evolution, the resulting retraining data do not necessarily provide representative coverage of the newly emerging traffic patterns. This raises the following question: why does detector-guided adaptation fail to fully recover classifier performance despite retraining on behaviorally non-conforming instances?

\vspace{-1mm}
\subsection{Uniform Traffic Sampling for Model Adaptation}
\label{subsec:uniformAdaptation}

\begin{figure}[!t]
    \centering
    \includegraphics[width=0.95\columnwidth]{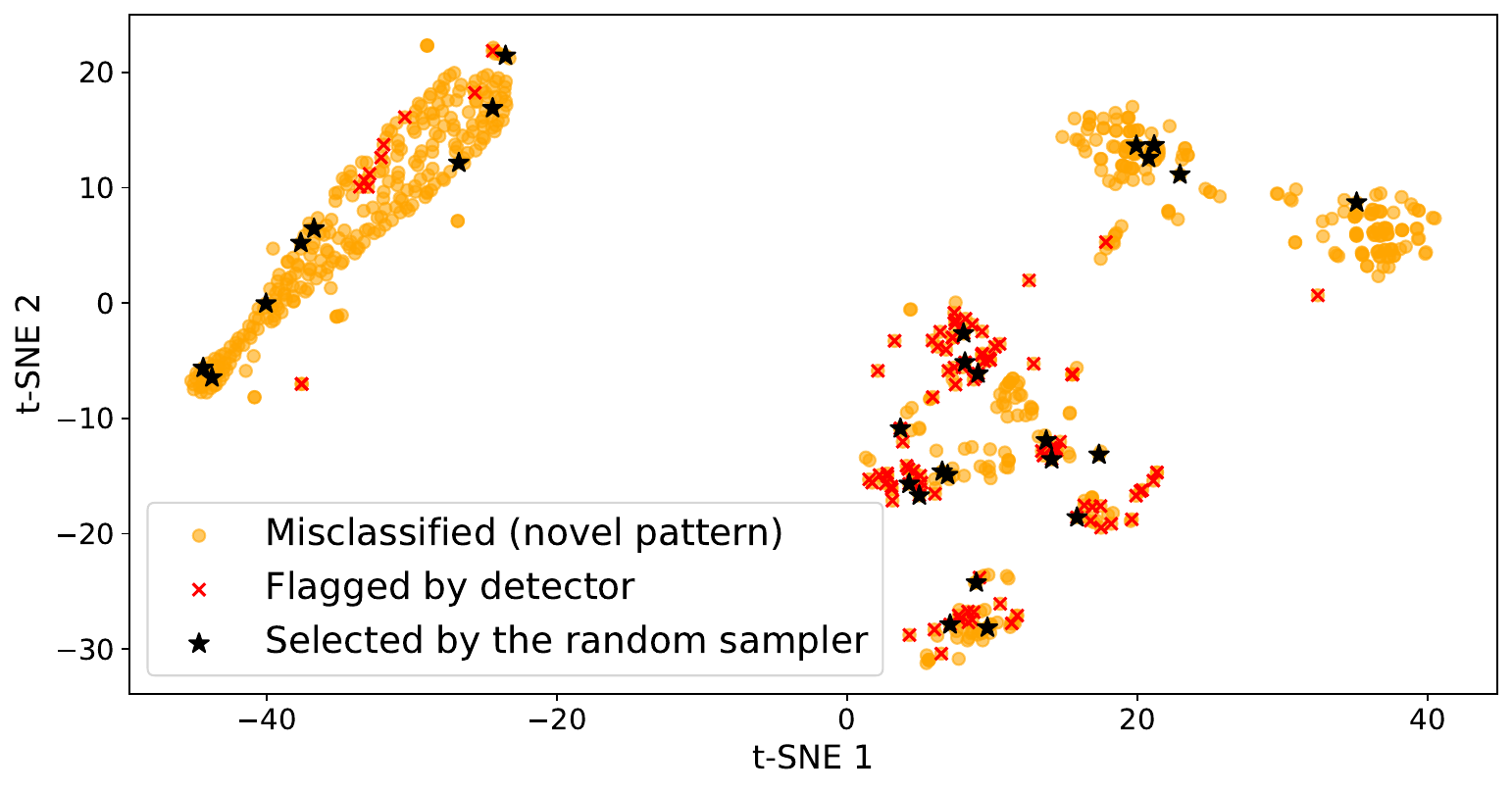}
    \caption{Detector-selected instances (red crosses) concentrate around only a subset of the emerging behavioral patterns (orange circles). Uniformly sampled instances (black stars) provide substantially broader coverage of the evolving traffic behavior, explaining their superior adaptation performance.}\label{fig:tSNE_random_selective}
    \vspace{-5mm}
\end{figure}

To answer the above question, we examine the deployment window starting from 30 July 2022 for the Google Nest Mini case study. 
Fig.~\ref{fig:tSNE_random_selective} compares the misclassified traffic instances in this window with those selected by the detector for retraining.  
The detector-selected instances concentrate on only a subset of the emerging behavioral patterns, leaving other newly formed regions unexplored. Consequently, retraining only on the detected instances fails to adequately capture the diversity of the evolving traffic behavior.

This observation motivates an alternative adaptation strategy. Rather than retraining on the detector-selected instances, we instead sample deployment traffic uniformly while matching the labeling rate of the detector. Fig.~\ref{fig:tSNE_random_selective} shows that uniformly sampled instances provide substantially broader coverage of the emerging behavioral patterns than the detector-selected samples.
 
We next evaluate this strategy over the entire deployment period.
One practical challenge is determining an appropriate sampling rate for each deployment window. 
To match the traffic labeling rate to the severity of the drift, we label instances at the same abundance as the detector flags at each step, and then update both the classifier and the detector using the newly labeled examples. The corresponding results are shown in Fig.~\ref{fig:classificationPerformance} (the dash-dotted curve {\color{black}with errorbars}). {\color{black} Note that throughout the remainder, we have repeated the evaluations involving random traffic sampling with 10 different random seeds, reporting the average $\pm$ the standard deviation of each metric over seeds.} The classification performance is consistently high and close to that of the exhaustive full adaptation approach. The window-average of F1-score is {\color{black}$0.9647 \pm 0.0003$}, and is achieved by labeling {\color{black}$10.15 \pm0.28\%$} of the aggregated deployment instances. The per-window labeling effort is depicted in Fig.~\ref{fig:labeling cost}.

Comparison of labeling costs of the two approaches in Fig.~\ref{fig:labeling cost} further explains the observed difference in adaptation performance.  
The number of detector-selected instances declines rapidly over time because progressively fewer instances are identified as behaviorally non-conforming. 
As a result, a considerable portion of the newly emerging behavioral patterns remain unlabeled and are therefore never incorporated into either the classifier or the behavioral model. Consequently, the detector gradually falls behind the evolving traffic behavior, causing the observed selection bias to persist over time.

\vspace{-1mm}
\subsection{Adaptation Under Operational Labeling Budgets}
Operational deployments typically impose strict limits on the amount of traffic that can be labeled. Having shown that uniformly sampled deployment traffic provides more effective classifier maintenance than detector-selected instances, we now investigate how the fixed labeling budget should be allocated over time. While the exact budget varies across operational environments, we use a representative labeling budget of $0.1\%$ of the deployment traffic per evaluation window, consistent with recent observations of practical traffic-labeling costs \cite{trainingWith1mille:CCS25}.

\begin{figure}[!t]
    \centering
    \includegraphics[width=0.95\columnwidth]{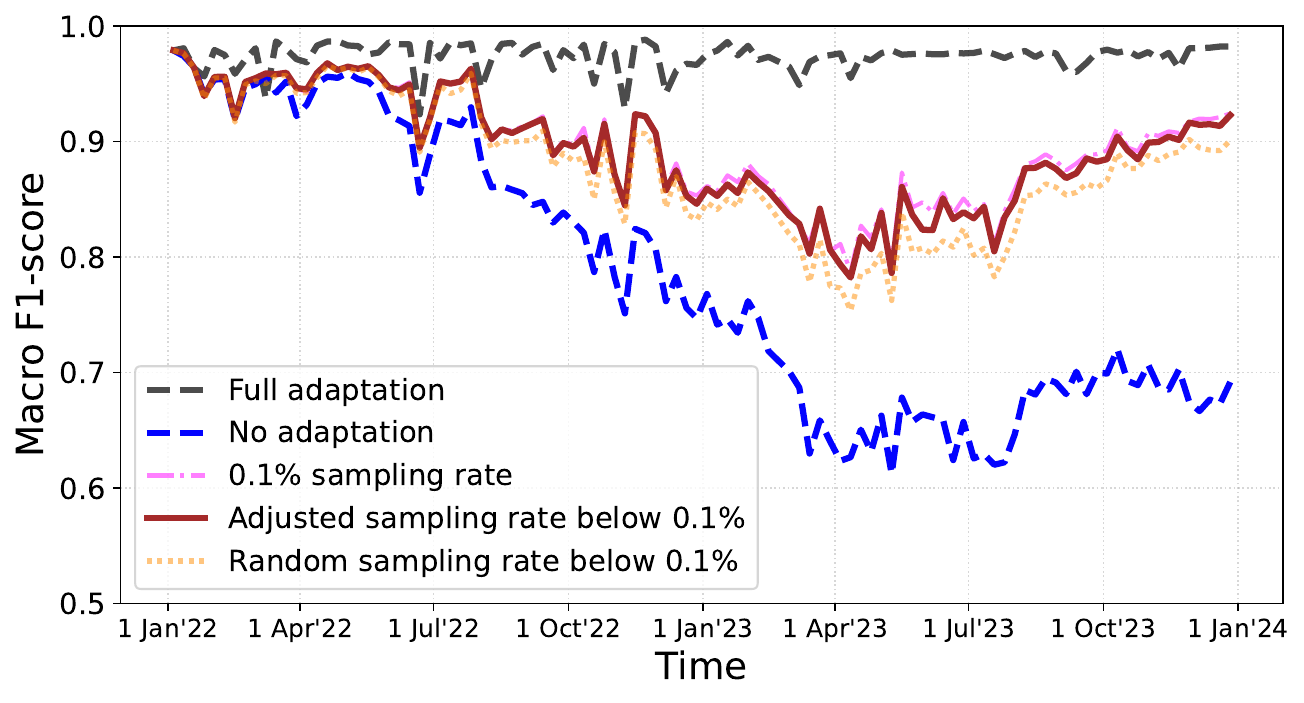}
    \caption{Classifier maintenance under a fixed operational labeling budget. Sampling uniformly at the full budget (dash-dotted pink) provides a reference operating point. Reducing the sampling rate uniformly across deployment (dotted orange) noticeably degrades adaptation performance. In contrast, adjusting the sampling rate according to the observed behavioral evolution (solid maroon) achieves {\color{black}comparable} performance while labeling fewer deployment instances. {\color{black}Error bars are omitted for visual clarity.}}
    \label{fig:adaptation_1mille}
    \vspace{-1mm}
\end{figure}

We first establish a reference operating point by uniformly sampling deployment traffic at the full budget of $0.1\%$ in every evaluation window. The corresponding adaptation performance is shown by the dash-dotted pink curve in Fig.~\ref{fig:adaptation_1mille}. Despite operating under a substantially tighter labeling budget, the resulting classifier maintains an average macro F1-score of {\color{black}$0.8970 \pm 0.0042$} across the deployment period.
However, this performance is achieved by fully utilizing the available labeling budget, raising the question of whether similar performance can be obtained while labeling even fewer deployment instances. A natural baseline is to reduce the labeling effort uniformly throughout deployment. To this end, we independently sample the labeling rate for each evaluation window from the uniform distribution $\mathrm{Unif}[0,\mathrm{Cap}]$, where $\mathrm{Cap}=0.1\%$. The resulting adaptation performance is shown by the dotted orange curve in Fig.~\ref{fig:adaptation_1mille}. {\color{black}Although this strategy reduces the aggregate labeling cost to $0.0501\pm0.0022\%$, the average F1-score across windows decreases to $0.8797\pm0.0067$, indicating a loss in adaptation performance when the labeling rate is reduced without accounting for behavioral evolution.}

These results indicate that the available labeling budget should be allocated in proportion to the observed level of behavioral evolution, rather than uniformly across deployment.
Specifically, for each evaluation window, we compute the fraction of deployment instances that are behaviorally non-conforming and record the maximum value observed so far. The labeling rate for the current window is then scaled linearly to this running maximum. Before moving to the next deployment window, both the classifier and detector are updated using the newly labeled instances. This procedure is outlined in Algorithm~\hyperlink{alg:driftAwareBelowBudget}{2}. Note that $\mathcal{R}(W;\Lambda)$ denotes randomly choosing from window $W$ at rate $\Lambda$. The resulting adaptation performance is shown by the solid maroon curve in Fig.~\ref{fig:adaptation_1mille}. {\color{black}Compared with uniform rate selection, adjusting the labeling rate according to the observed level of behavioral evolution improves classifier maintenance. Moreover, its average macro F1-score ($0.8934\pm0.0049$) remains close to full-budget sampling ($0.8970\pm0.0042$), while reducing the labeling cost from $0.1\%$ to $0.0826\pm0.0006\%$.}

\begin{algorithm}[t!]
\hypertarget{alg:driftAwareBelowBudget}{}
\caption{Budget-Aware Labeling Rate.}
\begin{algorithmic}
\STATE \textbf{Input:} A labeled training dataset $W^*_0$ and unlabeled deployment windows of traffic $\{W_i| 1\leq i\leq T\}$, sensitivity threshold $\epsilon$, maximum labeling budget of $C$.
\STATE \textbf{Output:} Progressively evaluated performance $\{\Pi_i| 1\leq i\leq T\}$, as well as detection labeling cost $\{\Lambda_i| 1\leq i\leq T\}$ in all deployment windows.
\STATE $W_{cum, 0}^* \leftarrow W^*_0$
\STATE $M_0 \leftarrow \mathcal{M}(W^*_0)$
\STATE $D_0 \leftarrow \mathcal{D}(W^*_0;\epsilon)$
\STATE $\Delta_{max} \leftarrow 0$
\FOR{$i = 1$ to $T$}
\STATE $\Pi_i \leftarrow \mathrm{perf}(W_i^*;M_{i-1})$ \hspace{0.0cm} {\color{gray} // In iteration $i$, ground-truth labels for $W_i$ are only available for the performance evaluation purpose.}
\STATE $\Delta_i \leftarrow |\tilde{D}_{i-1}(W_{i};M_{i-1})|$
\STATE $\Delta_{max} \leftarrow \max(\Delta_{max}, \Delta_i)$
\STATE $\Lambda_i \leftarrow \frac{\Delta_{i}}{\Delta_{max}} \times C$

\STATE $W^*_{cum,i}\leftarrow W^*_{cum,i-1}\parallel \mathcal{R}^*(W_{i};\Lambda_i)$ {\color{gray} // The $||$ operator denotes concatenation.}
\STATE $M_i \leftarrow \mathcal{M}(W^*_{cum,i})$
\STATE $D_i \leftarrow \mathcal{D}(W^*_{cum,i})$
\ENDFOR
\STATE \textbf{return} $\{\Pi_i| 1\leq i\leq T\}$, $\{\Lambda_i| 1\leq i\leq T\}$
\end{algorithmic}
\end{algorithm}

\setlength{\tabcolsep}{2pt}
\begin{table*}[!t]
\caption{Comparison of adaptation strategies under operational labeling constraints. Behavior-aware sampling achieves performance comparable to full-budget sampling while reducing the amount of labeled deployment traffic. Compared with confidence-based alternatives, it also provides feature-level interpretability.}

\label{tab:adaptationResults}
\centering
\resizebox{\textwidth}{!}{
    \renewcommand{\arraystretch}{1.1}
    \begin{tabular}{lllllll}
    \hline
    \textbf{Metric} & Cap. rate & Uniform below cap. & \textbf{Behavior-aware rate} & Low-confidence & {\color{black}Low-confidence stratified} & \textbf{Confidence-guided rate}\\
    \hline
    \textbf{Avg. macro F1-score} & {\color{black}$0.8970\pm0.0042$} & {\color{black}$0.8797 \pm 0.0067$} & \textbf{{\color{black}$0.8934 \pm 0.0049$}} & {\color{black}$0.8845$} & {\color{black}$0.8905$} & \textbf{{\color{black}$0.8942\pm0.0039$}}
    \\
    \textbf{Labeling cost} & $0.1\%\:\mathrm{(Cap)}$ & {\color{black}$0.0501\pm0.0022\%$} & \textbf{{\color{black}$0.0826\pm0.0006\%$}} & $0.1\%\:\mathrm{(Cap)}$ & $0.123\%$ & \textbf{{\color{black}$0.0867\pm0.0009\%$}} \\
    \hline
    \end{tabular}
}
\end{table*}

\subsection{Comparison to Confidence-based Model Adaptation}
\label{subsec:confidence-based}
Confidence-based active learning is a common strategy for adapting machine learning models under distribution shift, where deployment instances with the lowest prediction confidence are selected for labeling and retraining \cite{confidenceBasedActiveLearning}. 
Since our proposed method also provides a signal for guiding classifier maintenance, it is natural to compare the behavioral signal against classifier confidence.
{\color{black}A key limitation of confidence-based approaches is that they indicate prediction uncertainty without revealing which traffic features have changed. In contrast, the proposed behavioral signal identifies the non-conforming traffic features responsible for the detected behavioral evolution. We are therefore interested in assessing whether these interpretability benefits can be achieved without compromising adaptation performance.} We evaluate three confidence-based adaptation strategies under the same experimental setting. 

In the first experiment, we label $0.1\%$ of the instances with the lowest confidence in each window and use them to adapt the model. The resulting procedure yields an average macro F1-score of $0.8845$ during the deployment phase. {\color{black}This performance is lower than both full-budget uniform sampling ($0.8970$) and the behavior-aware strategy ($0.8934$).  It uses the same labeling budget as the former ($0.1\%$) and more labeled traffic than the latter ($0.0826\%$).}

We also evaluate a stratified version in which the labeling budget is applied independently to each predicted class.
Whenever the allocated budget would otherwise result in zero labeled instances for a class, we label one instance to ensure that every predicted class contributes to adaptation.
This approach results in an average F1-score of $0.8905$, albeit at a labeling cost of $0.123\%$. 
{\color{black}Despite the higher labeling cost, stratified low-confidence sampling achieves performance comparable to, but not higher than, the behavior-aware strategy.}
 
These experiments further support the observation that the adaptation signal is more effective for determining the traffic labeling rate than for direct instance selection. We therefore construct a counterpart to the behavior-aware strategy, modifying Algorithm~\hyperlink{alg:driftAwareBelowBudget}{2} by replacing the behavioral signal with classifier confidence while continuing to sample deployment instances uniformly. Specifically, we replace the drift severity signal (denoted by $\Delta_i$ in Algorithm~\hyperlink{alg:driftAwareBelowBudget}{2}) with the number of deployment instances whose prediction confidence falls below a predefined threshold. For this experiment, the threshold is set to a representative value of $80\%$.
The mentioned approach yields an average F1-score of {\color{black}$0.8942 \pm 0.0039$} over the deployment phase at a labeling cost of {\color{black}$0.0867 \pm 0.0009\%$}. {\color{black}The resulting macro F1-score matches that of the behavior-aware strategy while requiring a slightly higher labeling cost.} The performance of confidence-based methods, along with the behavior-aware strategy, over the entire deployment phase is depicted in Fig.~\ref{fig:comparison_to_confidence} in Appendix~\ref{app:confidence-based temporal}.

{\color{black}The quantitative comparison of all adaptation strategies is summarized in Table~\ref{tab:adaptationResults}. Overall, confidence-guided and behavior-guided sampling achieve comparable adaptation performance, while the former incurs a slightly higher labeling cost. Unlike classifier confidence, the behavioral signal identifies the traffic features responsible for behavioral evolution, providing feature-level interpretability while guiding the labeling rate.}

\subsection{Goodness Properties of the Detector and its Maintenance Scheme}\label{sec:goodness}
{\color{black}
The previous subsections demonstrated that  both the classifier and the detector are updated incrementally throughout deployment. 
Since the detector itself evolves over time, it is important to verify that this maintenance process preserves previously learned behavioral patterns while accurately incorporating newly emerging ones. Therefore, we evaluate the proposed detector and its maintenance scheme according to three desirable properties. These properties capture three complementary aspects of detector maintenance: preservation of previously learned behavior (\textit{stability}), robustness to selective updates (\textit{consistency}), and approximation of full-data updates (\textit{efficiency}).} 

\textbf{(P1) Stability}: Updating the detector with newly observed data should not introduce spurious detections (compared to the frozen detector) in subsequent windows, \ie previously learned patterns should be preserved. 

\textbf{(P2) Consistency}: The stability property should hold when the detector is updated using only the selected drifted instances instead of all labeled data. 

\textbf{(P3) Efficiency}: Updating the detector using only selected instances should produce detection behavior similar to updating with all labeled data, indicating that the underlying marginal distributions are adequately captured with significantly fewer labeled samples.

To evaluate these properties, consider the cumulative dataset $W^*_{cum,i-2}$ together with adjacent windows $W_{i-1}$ and $W_i$. Recall that, the $^*$ symbol denotes the labeled version corresponding to each dataset. We define three different detector update strategies before applying them to $W_i$:

\begin{equation}
    \text{No adaptation: }D_{ NA,\:i-1}=D_{i-2}=\mathcal{D}(W^*_{cum,i-2};\epsilon),
\end{equation}

\begin{equation}
    \text{Full: }D_{Full,\:i-1}=\mathcal{D}(W^*_{cum,i-2}\parallel W^*_{i-1};\epsilon),
\end{equation}

\begin{equation}
    \text{Selective: }D_{Sel,\:i-1}=\mathcal{D}\big(W^*_{cum,i-2}\parallel \tilde{D}^*_{i-2}(W_{i-1};M);\epsilon\big).
\end{equation}

A visualization of the three update strategies is provided in Appendix~\ref{app:detector_goodness}. Given the fixed classifier $M$ trained on the training data, we define the following detected instance sets: 
\begin{equation}
    S_{NA,i}=\tilde{D}_{NA,\:i-1}(W_i;M)
\end{equation}
\begin{equation}
    S_{Full,i}=\tilde{D}_{Full,\:i-1}(W_i;M)
\end{equation}
\begin{equation}
    S_{Sel,i}=\tilde{D}_{Sel,\:i-1}(W_i;M)
\end{equation}

If the three properties are satisfied, we expect: 
\begin{equation}
    S_{Full,i}\subseteq S_{NA,i}, \quad 
S_{Sel,i}\subseteq S_{NA,i}, \quad 
S_{Full,i}=S_{Sel,i}.
\end{equation}

Accordingly, we define the following metrics: 
\begin{equation}
    \eta^{(i)}_1 =\frac{|S_{NA,i}\cap S_{Full,i}|}{|S_{Full,i}|} \quad \text{(for stability)},
\end{equation}
\begin{equation}
    \eta^{(i)}_2 =\frac{|S_{NA,i}\cap S_{Sel,i}|}{|S_{Sel,i}|} \quad \text{(for consistency)},     
\end{equation}
\begin{equation}
    \eta^{(i)}_3 =\frac{|S_{Sel,i}\cap S_{Full,i}|}{|S_{Sel,i}\cup S_{Full,i}|} \quad \text{(for efficiency)}.
\end{equation}

{\color{black}
Across all deployment windows, the three metrics achieve average values of $\eta_1=0.984 \pm0.012$, $\eta_2=0.971\pm0.017$, and $\eta_3 =0.953\pm0.028$. These results indicate that incremental detector updates preserve previously learned behaviors, remain consistent when updated using only selected instances, and closely approximate updates using all labeled deployment traffic. Together, these observations support the proposed detector maintenance strategy for long-term deployment. Additional temporal results are provided in Appendix~\ref{app:detector_goodness}.}

%% file: sections/7limitations.tex
\vspace{-0.5mm}
{\color{black}
\section{Limitations}
Our study has several limitations. 
First, we infer behavioral evolution from marginal traffic features rather than their joint distribution. While this offers computational and data efficiency by avoiding joint distribution modeling, it is possible to overlook evolutions that occur only among the feature dependencies without impacting the marginals. As discussed in \S\ref{sec:detection}, this limitation is shared by input-based methods that avoid explicit modeling of high-dimensional joint feature distributions, which is generally prohibitive in practice. Second, our evaluation is based on a single deployment dataset collected by an ISP using a controlled testbed. Note that, unlike the existing public datasets that are collected over relatively short time frames, this dataset captures long-term behavioral evolution, which is central to the objectives of our work. Evaluating the proposed framework across additional deployment environments would further strengthen the generality of the conclusions.
Our focus in this paper has been on TCP/443 traffic (presumably HTTPS). Note that flow-level semantics vary across network protocols, being richer for hypertext protocols \cite{Arman:LCN}, and IoT devices are increasingly adopting TLS for network communications \cite{Paracha:IMC21}. Extending the proposed framework to additional protocols is an important direction for future work. 
Finally, our study is limited to Random Forest as the classification algorithm, which is well-established for traffic classification \cite{Yifan:IMC24}. 
Although our experiments employ Random Forests, the proposed detector uses only the predicted class and the corresponding class-conditional behavioral model. It therefore does not depend on the internal feature representation or confidence estimation mechanism of a particular classifier. Consequently, the proposed behavioral modeling framework can be combined with alternative traffic classifiers.
}

%% file: sections/8conclusion.tex
\section{Conclusion}
Accurate identification of IoT devices from network traffic is essential for security monitoring and policy enforcement.
However, as device behavior evolves over time, deployed classifiers experience concept drift, requiring periodic maintenance using newly labeled deployment traffic. 
In this paper, we first showed that behavioral evolution manifests differently across IoT device classes and that effective long-term classifier maintenance is feasible. We then developed a conformity-based detector that models class-conditional feature behavior directly from raw traffic features, providing interpretable feature-level explanations of behavioral evolution. Finally, we demonstrated that uniformly sampling deployment traffic while adjusting the labeling rate according to the observed behavioral evolution provides more effective and label-efficient classifier maintenance than selecting detector-identified instances alone.
These findings show that behavioral evolution is not merely a source of classifier degradation but also a practical signal for maintaining deployed IoT traffic classifiers under realistic operational constraints.

%% file: sections/apdx.tex
\subsection{Distribution Modeling Cases}
\label{app:distributionModelingCases}
Section~\ref{subsec:modelingFeatDistr} identified five possible cases when modeling feature distributions using KDE. This appendix summarizes those cases and their prevalence across the training data. To gain a better understanding of the composition of these scenarios, each one is denoted by the number of cases among classes and features that it accounts for across all class-feature pairs when modeling the training phase data. 
During detection, degenerate components are treated as perfectly conforming at their support points and non-conforming elsewhere, while continuous components are evaluated using the estimated density.

\begin{enumerate}
    \item (181 cases) Only a LNKDE component: when no training observation at $X=0$ exists and the dataset is not singular. Later in the test phase when performing detection, the PDF will be manually evaluated as $0$ at any realization of $X=0$, since $\log$ is not defined at $0$. 
    \item (4 cases) Only one degenerate component at $x_0>0$: when the training data is singular at $x_0$. Later in the test phase for detections, the PDF will be evaluated as ``infinity'' (to meet any threshold) at $X=x_0$ and $0$, elsewhere.
    \item (272 cases) One degenerate component at $X=0$ together with an LNKDE component: when the non-zero observations are not singular, allowing for the application of LNKDE. The PDF will be evaluated as ``infinity'' at $X=0$ and according to LNKDE, elsewhere.
    \item (1 case) Two degenerate components at $X=0$ and $X=x_0>0$: when the training data has realizations at $X=0$ and the rest of the datapoints are all equal to $x_0$. The PDF is evaluated as ``infinity'' at $X=0, x_0$ and $0$, elsewhere.
    \item (4 cases) One degenerate distribution at $X=0$: when the training samples are all equal to zero. The PDF will be evaluated as ``infinity'' at $X=0$ and zero, elsewhere.
\end{enumerate}

As expected, the vast majority of class-feature pairs are modeled using either LNKDE alone or LNKDE together with a point mass at zero, indicating that more complex cases occur only rarely in practice.

\begin{figure*}[!t]
    \centering
    \subfloat[Average drift indicator weights.\label{fig:indicator_weights}]{
        \includegraphics[width=0.90\columnwidth]{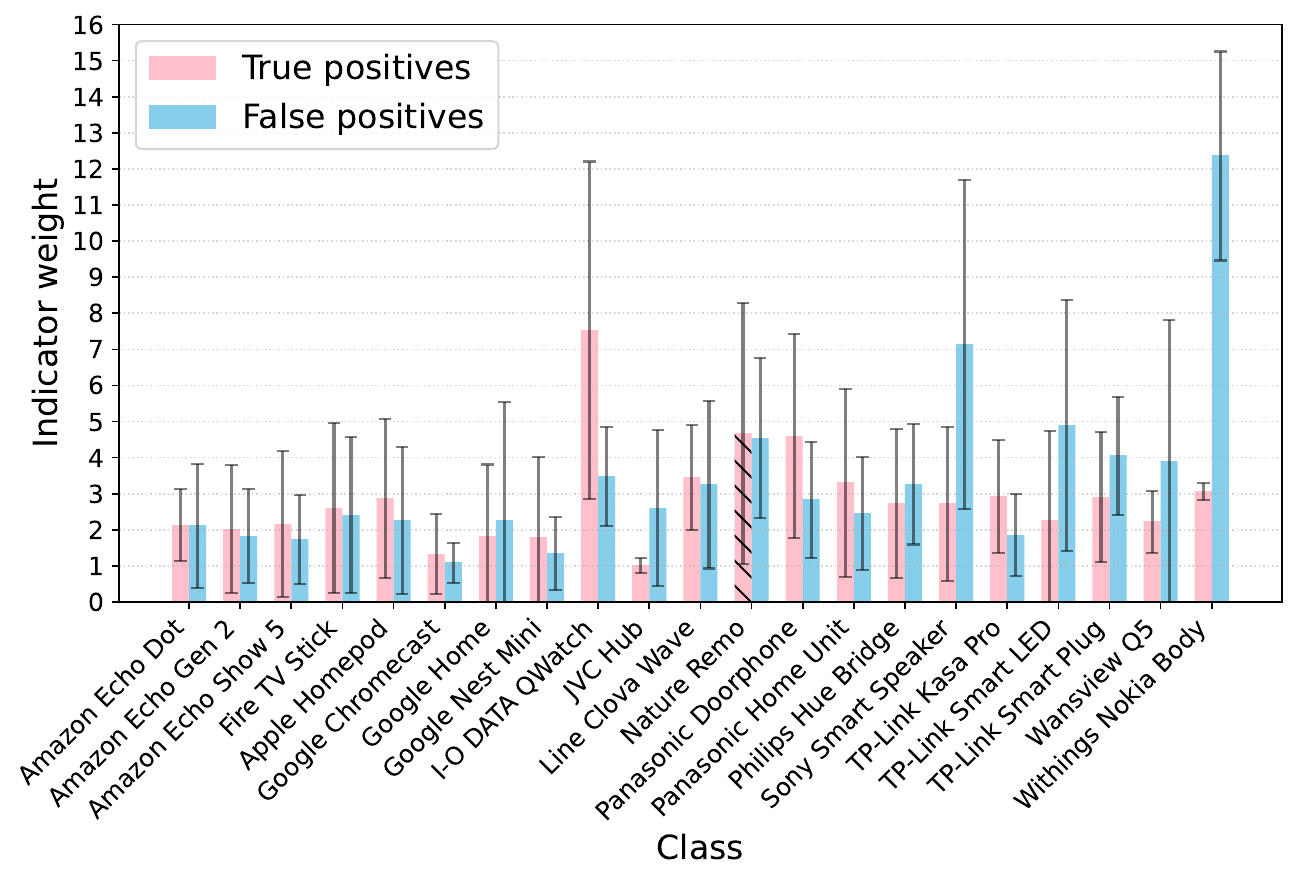}
    }
    \hfill
    \subfloat[Indicator entropy across classes and over time.\label{fig:indicator_entropy}]{
        \includegraphics[width=0.975\columnwidth]{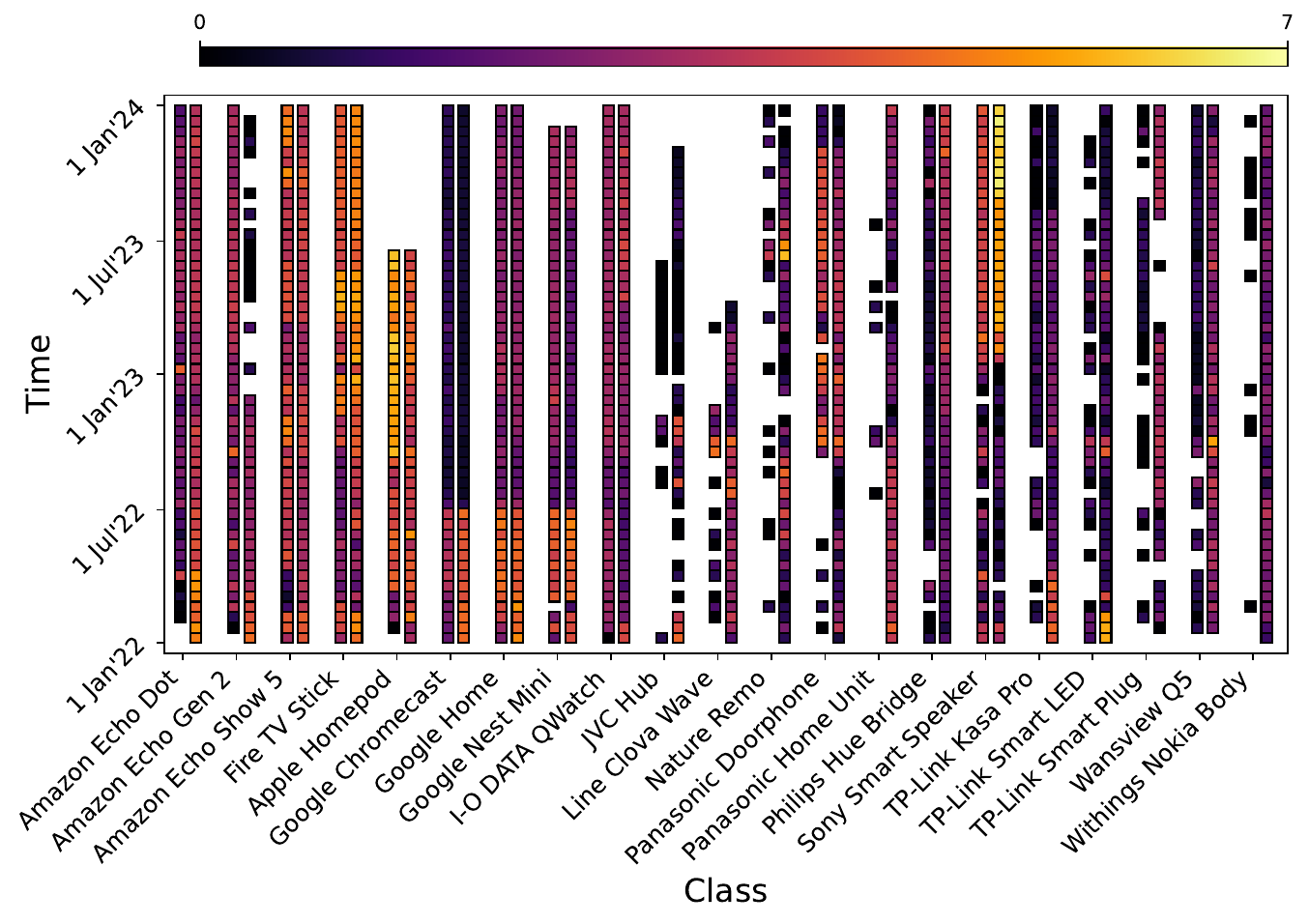}
    }
    \vspace{-1mm}
    \caption{Average drift indicator weights (a) show that the false positives and true positives cannot be effectively distinguished using the feature-drift indicator weights. The entropy trace (b) demonstrates a higher ambiguity in the indicators for false positives than true positives.}
    \label{fig:indicatorStats}
    \vspace{-4mm}
\end{figure*}



{\color{black}
\subsection{Sensitivity Analysis for $\epsilon$}
\label{app:sensitivityEpsilon}
We used $\epsilon=0.01$ throughout our experiments and evaluations presented in this paper. Here, we assess the sensitivity of the detector to values of $\epsilon$ in the neighborhood of this operating point. Table~\ref{tab:sensitivityEpsilon} summarizes the TPR, specificity and balanced accuracy results for values of $\epsilon$ near $0.01$. The results show relatively low sensitivity to the precise choice of $\epsilon$. Therefore, once $\epsilon$ has been selected at the appropriate order of magnitude (\eg $10^{-2}$), further fine-tuning has little impact on detector performance. 

\begin{table}[t]
\caption{Detection results show low sensitivity to $\epsilon$.}
\label{tab:sensitivityEpsilon}
\centering
\renewcommand{\arraystretch}{1.1}
\begin{tabular}{|c|c|c|c|}
\hline
\textbf{$\epsilon$} & \textbf{TPR} & \textbf{Specificity} & \textbf{Balanced accuracy} \\
\hline
$0.0050$ & $0.72\pm 0.23$ & $0.53\pm 0.29$ & $0.62\pm 0.16$ \\
$0.0075$ & $0.74\pm 0.23$ & $0.52\pm 0.28$ & $0.63\pm 0.16$ \\
$0.0100$ & $0.77\pm 0.22$ & $0.49\pm 0.27$ & $0.63\pm 0.15$ \\
$0.0125$ & $0.79\pm0.21$ & $0.48\pm0.26$ &$0.63\pm0.14$ \\
$0.0150$ & $0.80\pm0.20$ & $0.46\pm0.25$ &$0.63\pm0.13$ \\
\hline
\end{tabular}
\end{table}}

\balance
\subsection{Consistency of Feature-Level Explanations}\label{app:explainMore}

\subsubsection{Indicator Weights}
We first investigate whether true and false detections differ in the number of non-conforming features contributing to their explanation. Fig.~\ref{fig:indicator_weights} compares the indicator weights of true and false detections across device classes.  
No consistent separation is observed between the two groups, and substantial temporal variation is evident in both. These results indicate that the number of non-conforming features alone is insufficient to distinguish true detections from false ones.

\subsubsection{Diversity of Feature-Level Explanations}
We next investigate the feature-level explanations across classes over time for both the true and false positives. 
To quantify the diversity of feature-level explanations, 
we compute the entropy of the indicator bit strings within fortnightly windows for every device class. 
Fig.~\ref{fig:indicator_entropy} shows that true detections (left column) consistently exhibit lower entropy than false detections (right column), indicating that true behavioral evolution tends to produce more structured and repeatable explanation patterns.

Overall, the indicator-weight and entropy analyses show that feature-level explanations remain structured across deployment, while true behavioral evolution exhibits considerably lower explanation diversity than false detections. These observations further support the detector's ability to provide meaningful and interpretable explanations of behavioral evolution.

\begin{figure}[H]
    \centering
    \includegraphics[width=0.95\columnwidth]{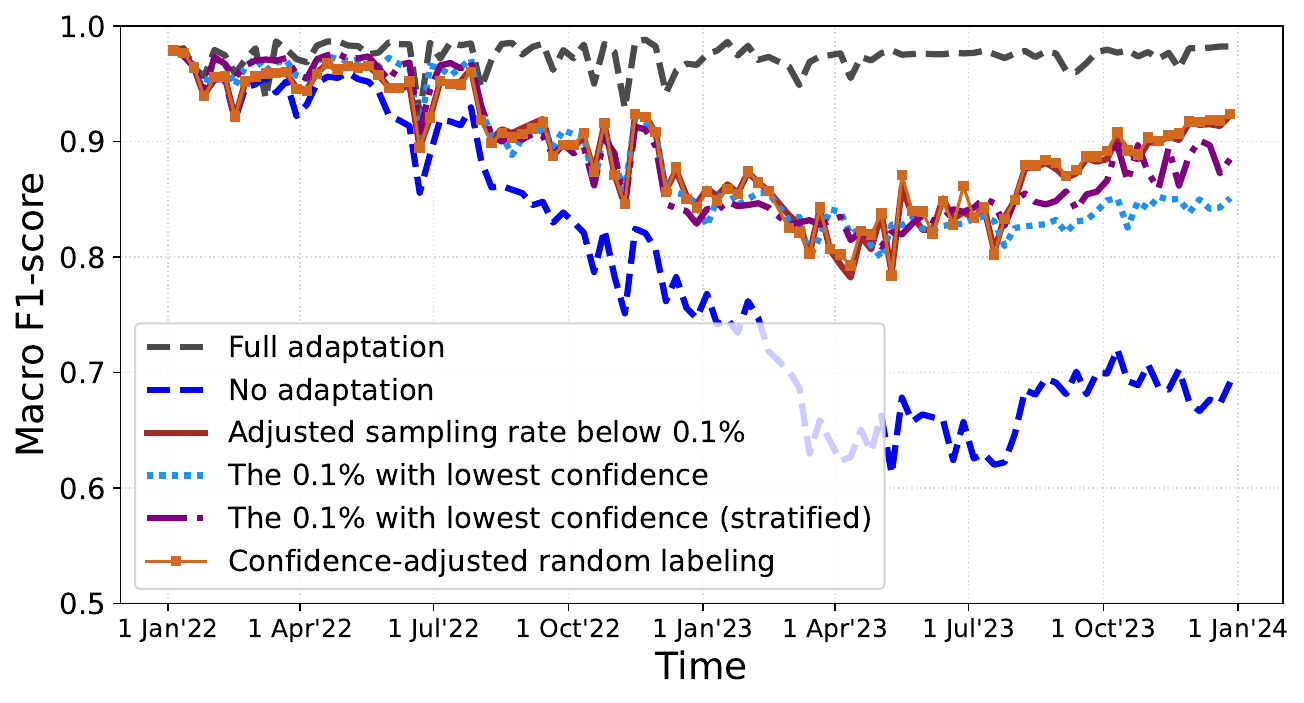}
    \caption{Temporal comparison of confidence-based and behavior-aware adaptation strategies. }
    \label{fig:comparison_to_confidence}
    \vspace{-1mm}
\end{figure}

\subsection{Formal Description of Detector-guided Adaptation}
\label{app:detector-guidedAdapt}
The formal description of the detector-guided adaptation scheme (as discussed in \S\ref{subsec:detector-guidedAdapt}) is presented in Algorithm~\hyperlink{alg:adaptwithdetect}{3}.

\begin{algorithm}[t!]
\hypertarget{alg:adaptwithdetect}{}
\caption{Adaptation Using the Detected Instances.}
\begin{algorithmic}
\STATE \textbf{Input:} A labeled training dataset $W^*_0$ and unlabeled deployment windows of traffic $\{W_i| 1\leq i\leq T\}$, sensitivity threshold $\epsilon$.
\STATE \textbf{Output:} Progressively evaluated performance $\{\Pi_i| 1\leq i\leq T\}$, as well as detection labeling cost $\{\Lambda_i| 1\leq i\leq T\}$ in all deployment windows.
\STATE $W_{cum, 0}^* \leftarrow W^*_0$
\STATE $M_0 \leftarrow \mathcal{M}(W^*_0)$
\STATE $D_0 \leftarrow \mathcal{D}(W^*_0;\epsilon)$
\FOR{$i = 1$ to $T$}
\STATE $\Pi_i \leftarrow \mathrm{perf}(W_i^*;M_{i-1})$ \hspace{0.0cm} {\color{gray} // In iteration $i$, ground-truth labels for $W_i$ are only available for the performance evaluation purpose.}
\STATE $\tilde{D}^*_{i-1}(W_{i};M_{i-1}) \leftarrow \mathrm{label}(\tilde{D}_{i-1}(W_{i};M_{i-1}))$
\STATE $\Lambda_i \leftarrow |\tilde{D}_{i-1}(W_{i};M_{i-1})|$
\STATE $W^*_{cum,i}\leftarrow W^*_{cum,i-1}\parallel \tilde{D}^*_{i-1}(W_{i};M_{i-1})$ {\color{gray} // The $||$ operator denotes concatenation.}
\STATE $M_i \leftarrow \mathcal{M}(W^*_{cum,i})$
\STATE $D_i \leftarrow \mathcal{D}(W^*_{cum,i})$
\ENDFOR
\STATE \textbf{return} $\{\Pi_i| 1\leq i\leq T\}$, $\{\Lambda_i| 1\leq i\leq T\}$
\end{algorithmic}
\end{algorithm}

\subsection{Performance of the Confidence-based Methods Over Time}
\label{app:confidence-based temporal}
{\color{black}
Fig.~\ref{fig:comparison_to_confidence} presents the temporal classification performance of the behavior-aware and confidence-guided sampling strategies, together with two confidence-based instance-selection baselines discussed in \S\ref{subsec:confidence-based}. The behavior-aware and confidence-guided strategies achieve comparable performance throughout deployment and generally outperform the two confidence-based instance-selection approaches. As reported in Table~\ref{tab:adaptationResults}, the behavior-aware strategy achieves this performance at a slightly lower labeling cost while additionally offering feature-level explanations of behavioral evolution.
}



\begin{figure*}[!t]
    \centering
    \subfloat[Detector update strategies for evaluating stability, consistency, and efficiency.\label{fig:detectorEvaluationScheme}]{
        \includegraphics[width=0.90\columnwidth]{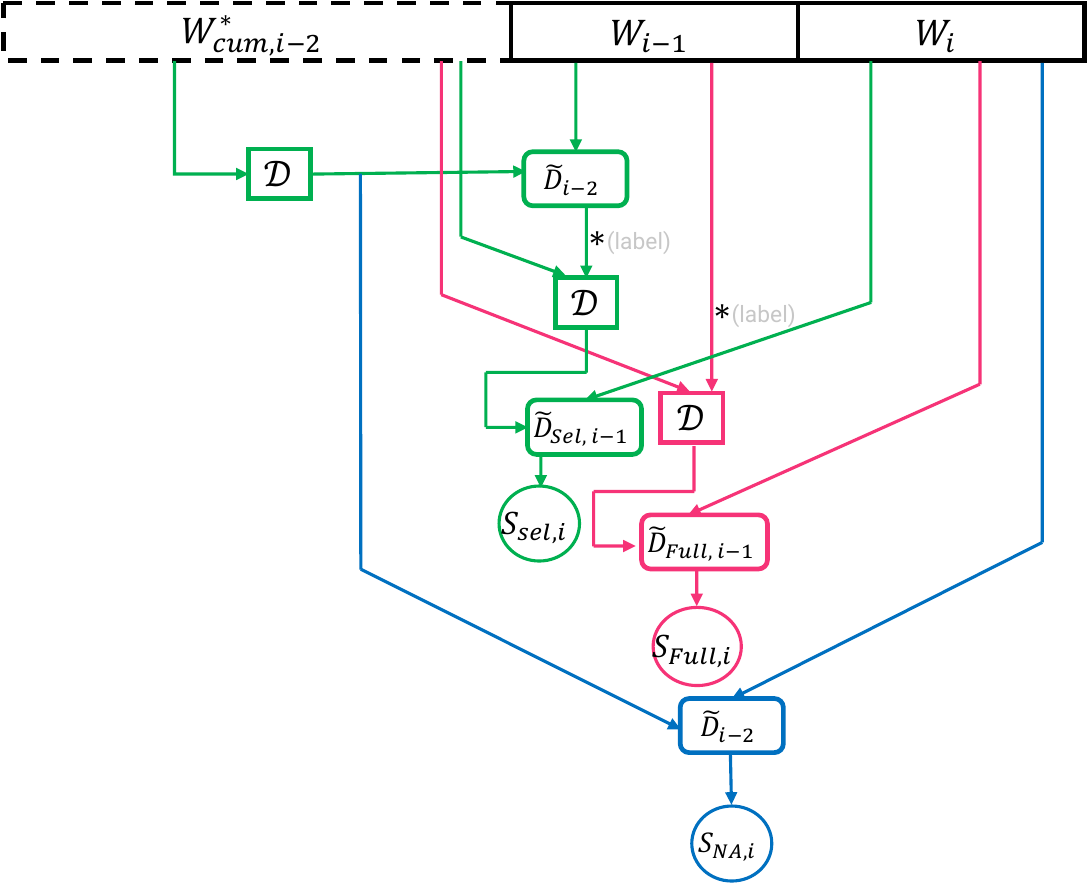}
    }
    \hfill
    \subfloat[Evaluation metrics ($\eta_1$, $\eta_2$, and $\eta_3$) over the deployment period.\label{fig:goodnessMetrics}]{
        \includegraphics[width=0.975\columnwidth]{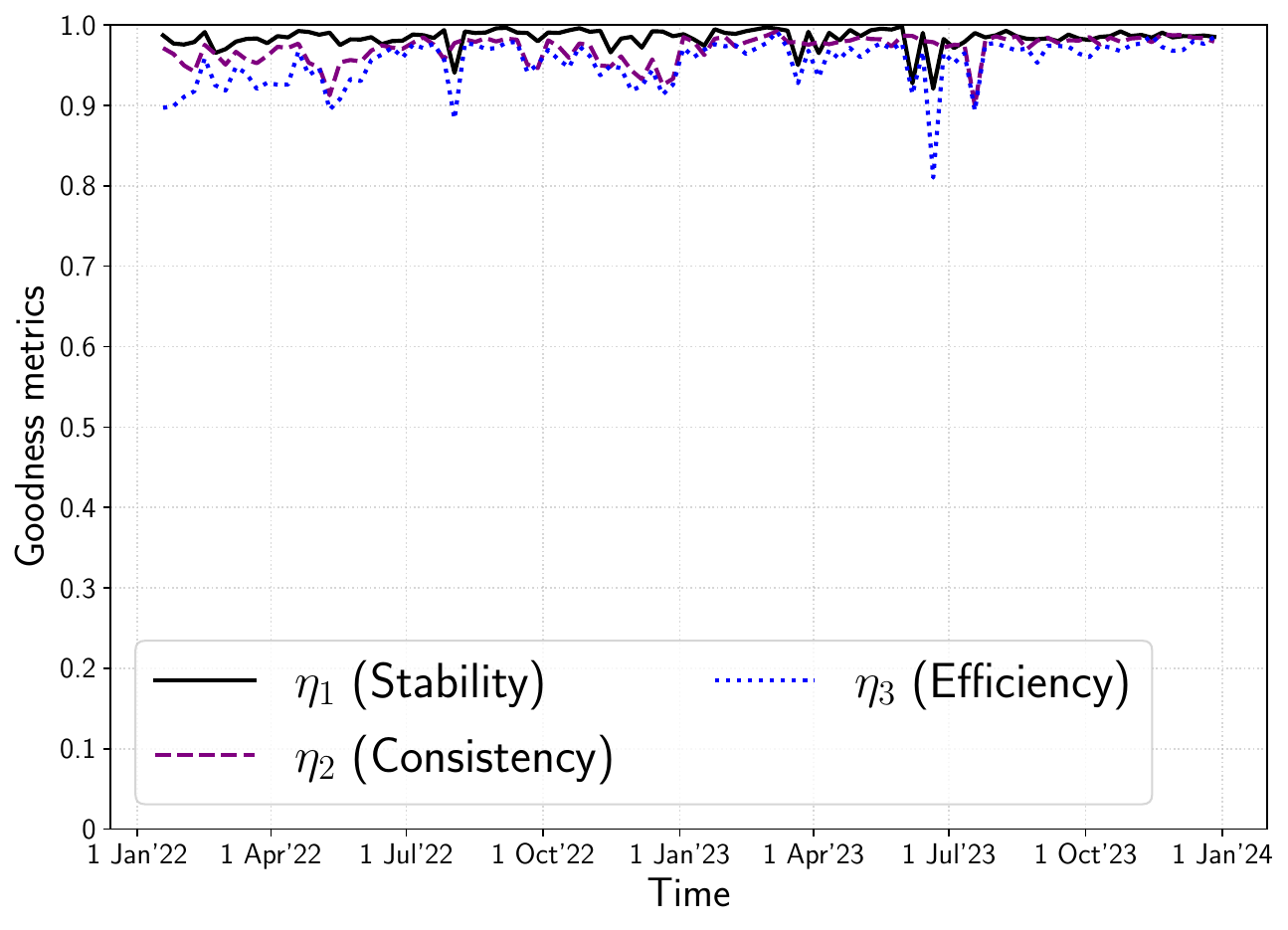}
    }
    \vspace{-1mm}
    \caption{Evaluation of the behavioral signal for adaptation. (a) Detector update strategies used to assess stability, consistency, and efficiency. (b) Corresponding evaluation metrics over the deployment period. 
    Values close to one indicate that detector updates using only selected instances closely preserve the behavior of updates performed using all labeled deployment traffic.}
    \label{fig:measuringGoodness}
    \vspace{-4mm}
\end{figure*}

{\color{black}
\subsection{Measures of Detector Goodness for Adaptation}
\label{app:detector_goodness}
This appendix provides the temporal results supporting the summary statistics reported in \S\ref{sec:goodness}.
As discussed in \S\ref{sec:adaptEval}, we evaluate the detector according to three desirable properties: stability ($\eta_1$), consistency ($\eta_2$), and efficiency ($\eta_3$). Computation of the goodness metrics relies on detection results performed continuously based on three different strategies as demonstrated in Fig.~\ref{fig:detectorEvaluationScheme}. The classifier model $M$ is fitted on the training data and kept fixed throughout the process. The goodness metrics $\eta_1$, $\eta_2$ and $\eta_3$ are depicted over time in Fig.~\ref{fig:goodnessMetrics}. The results show that all three properties are consistently satisfied throughout deployment.}